\documentclass[fleqn,usenatbib]{mnras}
\usepackage{newtxtext,newtxmath}
\usepackage[T1]{fontenc}
\DeclareRobustCommand{\VAN}[3]{#2}
\let\VANthebibliography\thebibliography
\def\thebibliography{\DeclareRobustCommand{\VAN}[3]{##3}\VANthebibliography}

\usepackage[utf8]{inputenc}
\usepackage[figuresright]{rotating}%只旋转表格中某个单元格的标题
\usepackage{multirow}%跨行表格宏包
\usepackage{array}%数组和表格制作
\pdfoutput=1 %for arXiv submission，将输出格式设置为 PDF
\usepackage{amsmath,amstext}%多种公式环境和数学命令，在公式环境使用\text
\usepackage{CJK}%中文处理支持

\usepackage[T1]{fontenc}%使用其他编码字符
\usepackage[figure,figure*]{hypcap}%插图https://www.overleaf.com/project/62e9dbf3b389ca6ab455fd74
\usepackage{booktabs}%绘制水平表格线
\usepackage{paralist}%多种列表环境
\usepackage{natbib}%文献格式宏包
\usepackage{graphicx}%插图处理
\usepackage{float}
\usepackage{hyperref} %to use the \url or \href tag，创建超文本链接和PDF书签
\hypersetup{
colorlinks=true,
linkcolor=blue,
anchorcolor=blue,
citecolor=blue,
urlcolor=blue}%设置参考连接格式
\usepackage{cleveref}
\usepackage{indentfirst}%首段缩进
\graphicspath{{fig/}}
\usepackage{subfigure}
\usepackage{threeparttable}
\usepackage{siunitx}
\title[NS asymmetries and phase spiral]{North-South asymmetries in the Galactic thin disk associated with the vertical phase spiral as seen using LAMOST-{\it Gaia} stars}

\author[Lin et al.]{
Jun Lin$^{1}$,
Rui Guo$^{2}$,
Sarah A. Bird$^{1}$,
Haijun Tian$^{3,4,1,5}$\thanks{Email: hjtian@lamost.org, guorui20@sjtu.edu.cn, sarahbird@ctgu.edu.cn, liuchao@nao.cas.cn.},
Chao Liu$^{6}$,
Chris Flynn$^{7,8}$,
Gaochao Liu$^{1}$,
\newauthor and Sheng Cui$^{1}$
\\
$^{1}$Center for Astronomy and Space Sciences, China Three Gorges University, Yichang 443002, People's Republic of China.\\
$^{2}$Department of Astronomy, School of Physics and Astronomy, Shanghai Jiao Tong University, 800 Dongchuan Road, Shanghai 200240, People's Republic of China.\\
$^{3}$School of Science, Hangzhou Dianzi University, Hangzhou, 310018, People's Republic of China.\\
$^{4}$Space Information Research Institute, Hangzhou Dianzi University, Hangzhou, 310018, People's Republic of China.\\
$^{5}$School of Physics and Astronomy, China West Normal University, NanChong 637002, People's Republic of China.\\
$^{6}$Key Lab of Space Astronomy and Technology, National Astronomical Observatories, Chinese Academy of Sciences, 20A Datun Road,\\ Chaoyang District, Beijing 100101, People's Republic of China.\\
$^{7}$Center for Astrophysics and Supercomputing, Swinburne University of Technology, Post Office Box 218, Hawthorn, VIC 3122, Australia.\\
$^{8}$ARC Centre of Excellence for Gravitational Wave Discovery (OzGrav), Australia.
}

\date{Accepted XXX. Received YYY; in original form zzz}
\pubyear{2023}

\begin{document}
\label{firstpage}
\pagerange{\pageref{firstpage}--\pageref{lastpage}}
\maketitle

\begin{abstract}  
We select 1,052,469 (754,635) thin disk stars from {\it Gaia} eDR3 and LAMOST DR7 in the range of Galactocentric radius $R$ (guiding center radius $R_\mathrm{g}$) from 8 to 11\,kpc to investigate the asymmetries between the North and South of the disk midplane. More specifically we analyze the vertical velocity dispersion profiles ($\sigma_{v_{z}}(z$)) in different bins of $R$ ($R_\mathrm{g}$) and $[\mathrm{Fe/H}]$. We find troughs in the profiles of $\sigma_{v_{z}}(z)$ located in both the North ($z \sim 0.7$\,kpc) and South ($z \sim -0.5$\,kpc) of the disk at all radial and chemical bins studied. The difference between the Northern and Southern vertical velocity dispersion profiles ($\Delta\sigma_{v_{z}}(|z|)$) shows a shift between curves of different $R$ and $R_\mathrm{g}$. A similar shift exists in these NS asymmetry profiles further divided into different $[\mathrm{Fe/H}]$ ranges. The sample binned with $R_\mathrm{g}$ more clearly displays the features in the velocity dispersion profiles. The shift in the peaks of the $\Delta\sigma_{v_{z}}$ profiles and the variation in the phase spiral shape binned by metallicity indicate the variation of the vertical potential profiles and the radial metallicity gradient. The wave-like signal in NS asymmetry of $\sigma_{v_{z}}(z)$ largely originates from phase spiral; while the NS asymmetry profiles of [Fe/H] only display a weak wave-like feature near solar radius. We perform a test particle simulation to qualitatively reproduce the observed results. A quantitative explanation of the NS asymmetry in the metallicity profile needs careful consideration of the spiral shape and the perturbation model, and we leave this for future work.
\end{abstract}


\section{Introduction}
\par\indent The vertical density and kinematics of the Milky Way play an important role in understanding the structure and dynamics of the Milky Way as a whole. Data from large-scale sky surveys, e.g., Large Sky
Area Multi-Object Fiber Spectroscopic Telescope survey \citep[LAMOST,][]{Cui2012,Yan2022}, {\it Gaia} \citep{2021A&A...649A...1G}, Sloan Digital Sky Survey \citep[SDSS,][]{Almeida2023}, RAdial Velocity Experiment \citep[RAVE,][]{Steinmetz2006}, Gaia-ESO \citep{Gilmore2012,Randich2013}, GALactic Archaeology with HERMES \citep[GALAH,][]{De_Silva2015,Buder2021}, are vital to reveal the complex details of the vertical structure of the Milky Way; these data break the simple assumption of a North-South (NS) symmetric Milky Way disk in dynamical equilibrium. 

\citet{2012ApJ...750L..41W} found that the disk has a wave-like NS asymmetry in the star counts based on SDSS data, and the disk has a non-zero vertical bulk motion that slowly changes with vertical height. After carefully considering the influence of observational errors and selection biases, \citet{2013ApJ...777...91Y} confirmed the existence of a NS asymmetry in the stellar number counts. With the large sample size and high-precision parallaxes from {\it Gaia} DR2, \citet{2019MNRAS.482.1417B} found that the asymmetry in number density is independent of the color of main sequence stars. \citet{2022A&A...668A..95W} used {\it Gaia} DR3 data to make non-parametric estimates of the number density and the average velocity field of the disk. They found the results significantly deviate from the axisymmetric and NS mirror symmetric models. 

Several early studies finding asymmetries in the vertical bulk motion of the disk include. Many studies of the vertical bulk motion have found that there are two modes, i.e, the breathing mode and bending mode. The vertical bulk motion of the disk is a combination of these two modes \citep{2012ApJ...750L..41W,2013ApJ...777L...5C,2013MNRAS.436..101W,2014MNRAS.440.1971W,2018MNRAS.477.2858W,2020A&A...634A..66L}. \citet{Ding2021} selected K giant stars from LAMOST and studied the vertical kinematic structure in the Galactocentric radial range of $R\in 5-15$\,kpc and up to 3\,kpc vertically from the Galactic plane. They found that the vertical velocity dispersion in the South is larger than that in the North in the radial range of $7-13$\,kpc.

\citet{Antoja2018} found a spiral feature in the phase space of $z-v_z$ color-coded by either the azimuthal velocity ($v_\phi$) or the radial velocity ($v_{R}$), which indicates that the disk is undergoing a phase mixing process possibly initiated by a vertical perturbation. 

Besides the asymmetry in kinematics and density, the vertical distribution of metallicity in the Solar Neighbourhood is found to show a wave-like asymmetry, which  coincides with the previously known asymmetry in the stellar number density distribution \citep{2019ApJ...878L..31A}. In addition, the phase spiral structures in the density and kinematic distributions were found to be linked to the metallicity ([M/H]) variations \citep{2022arXiv220605534G}.
\begin{figure*}[]
    \centering
    \includegraphics[width=4.7cm,trim=0.1cm 0.2cm 0.2cm 0.13cm, clip]{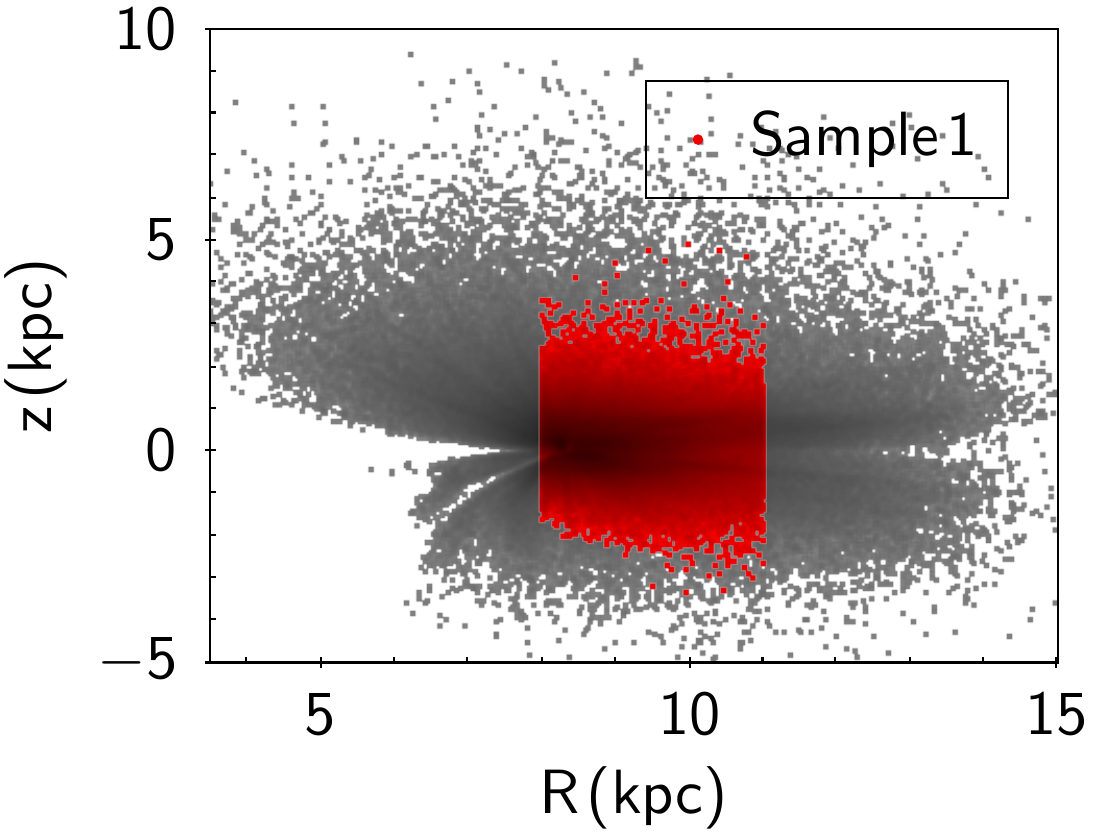}
    \includegraphics[width=3.9cm, trim=3.2cm 0.2cm 0.2cm 0.13cm, clip]{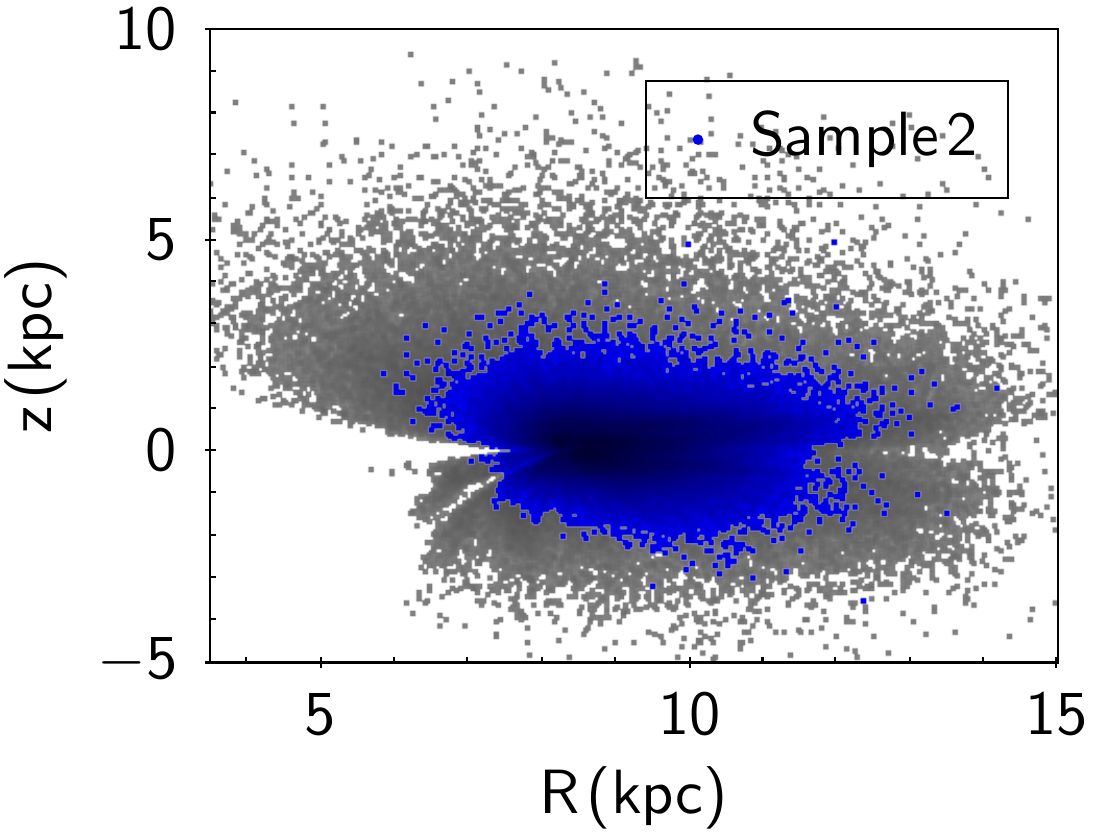}
    \includegraphics[width=4.7cm,trim=0.1cm 0.2cm 0.53cm 0.1cm, clip]{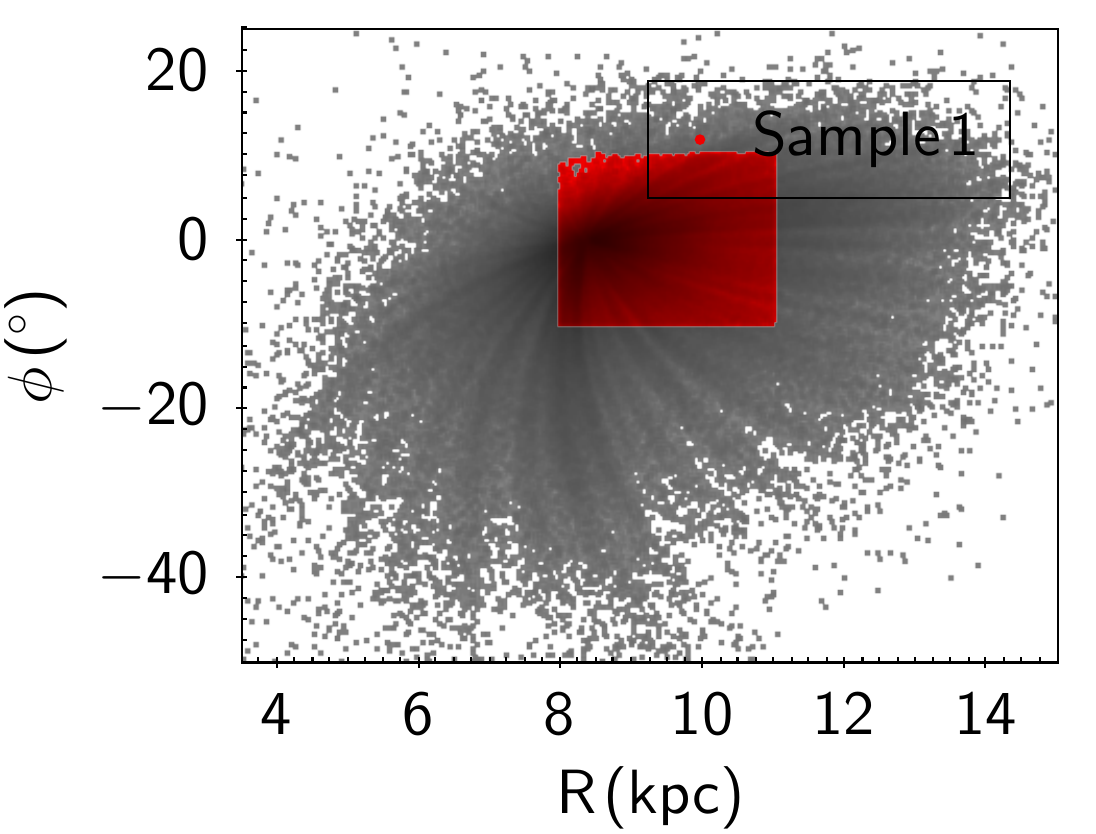}
    \includegraphics[width=3.9cm, trim=3.6cm 0.2cm 0.0cm 0.1cm, clip]{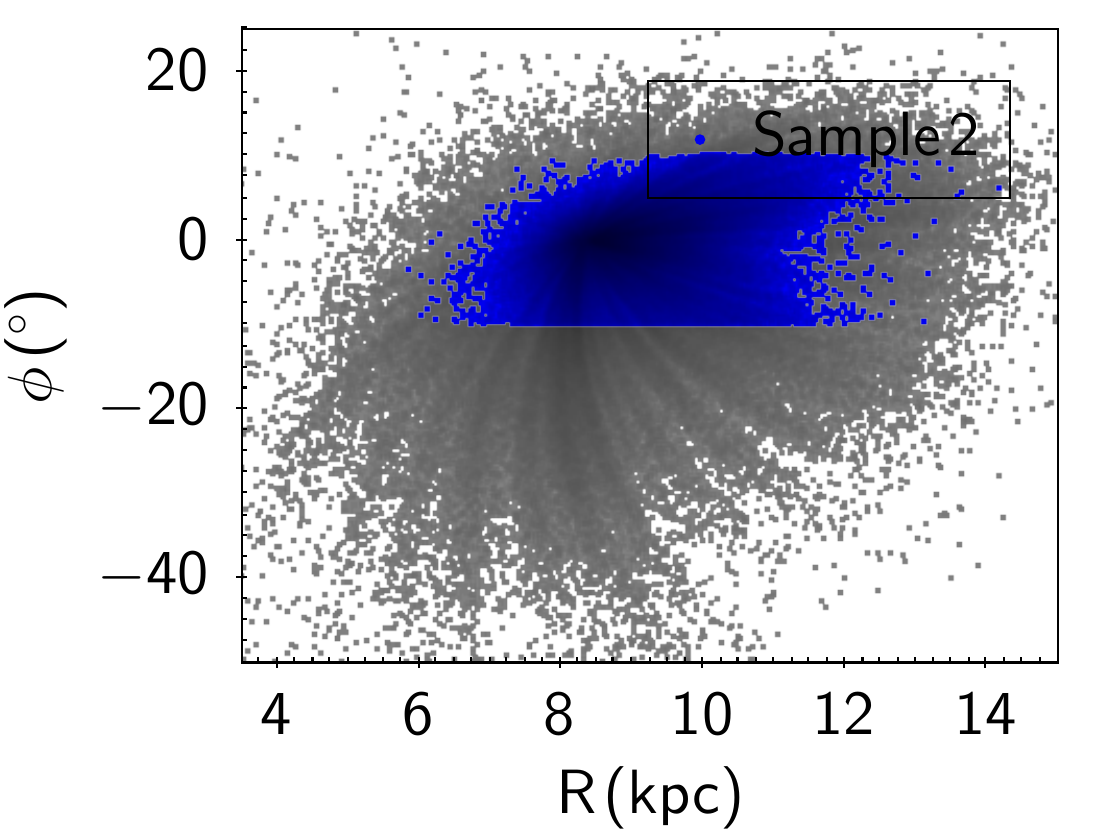}
    \caption{Distributions of our samples in Galactocentric cylindrical coordinates.  The first and second panels on the left show the distributions of Sample\,1 and Sample\,2 in $R-z$, respectively. The third and fourth panels show the distributions of Sample\,1 and Sample\,2 in $R-\phi$, respectively. The grey markers represent the catalog only after implementing cuts for quality. The red and blue markers represent Sample\,1 and Sample\,2, respectively. The data selection is detailed in Section \ref{data}. }
    \label{fig1}
\end{figure*}
\begin{figure}
    \centering
    \includegraphics[width=8cm, trim=0.3cm 1.5cm 2.0cm 2.5cm, clip]{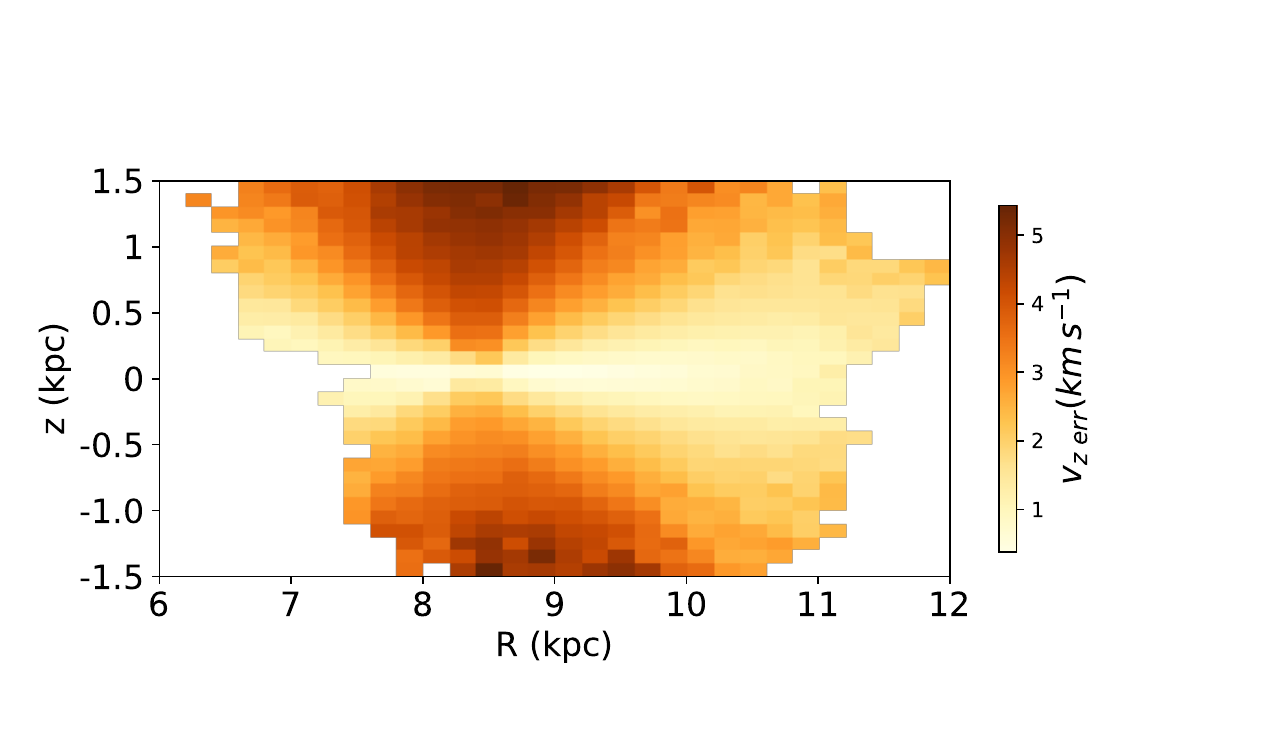}
    \includegraphics[width=8cm, trim=0.3cm 1.5cm 2.0cm 2.5cm, clip]{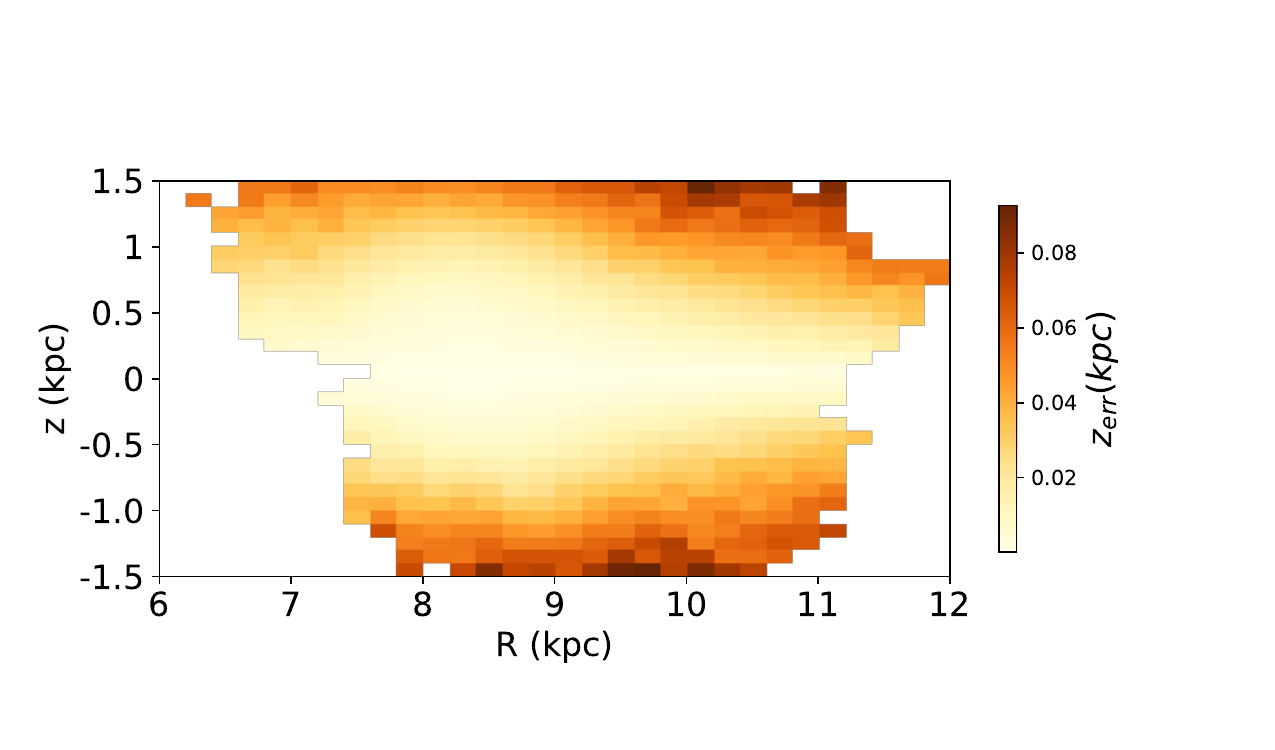}
    \caption{Median vertical velocity error (upper) and median $z$ error (bottom) distribution of LAMOST DR7 and {\it Gaia} eDR3 matched catalog after our implemented cuts including quality (listed in Section \ref{data}), $[\mathrm{Fe/H}]>-0.4$, and $|V-V_\mathrm{LSR}|<200$ km\,s$^{-1}$.}
    \label{fig2}
\end{figure}

In modelling our Galaxy, often the vertical velocity dispersion profile ($\sigma_{v_{z}}(z$)) is needed, such as in using the vertical Jeans equation to estimate the vertical force and local surface density and dark matter density \citep[e.g.,][]{2012ApJ...750L..41W,Hagen2018,2020A&A...643A..75S,Guo2020}. Understanding any asymmetries in ($\sigma_{v_{z}}(z$)) is essential as this will affect the derived parameters from the modelling \citep[e.g.,][]{Banik2017,2018MNRAS.478.1677S,Haines2019}.
\citet{2020A&A...643A..75S}'s Jeans analysis shows that velocity dispersion is more perturbed in the North than in the South of the disk midplane. They additionally find a connection between the NS asymmetry in number density and the vertical phase spiral. \citet{Guo2020} found the velocity dispersion curve shows a plateau feature at different heights in the North and South. These plateaux result in different dark matter density estimates between the North and South.

The vertical velocity dispersion distribution can reflect the history of dynamical heating, as characterised by the age-velocity dispersion relationship. The small scale irregular gravitational potential generated by giant molecular clouds and the transient spiral arm structures in the disk are considered to be important heating mechanisms \citep{1953ApJ...118..106S, 1984MNRAS.208..687L, 1967ApJ...150..461B, 10.1093/mnras/stw777, 2016MNRAS.462.1697A, 2019ApJ...878...21T}. External perturbation is also an important source of heating \citep[e.g.,][]{Gomez2012,2016MNRAS.459..199G}. Such perturbation can cause the bending modes of vertical bulk motion, and initiate the formation of a phase spiral in the $z-v_z$ phase space \citep{Antoja2018,Tiann2018ApJ,Binney2018,Bland-Hawthorn2019,2019MNRAS.485.3134L,2021MNRAS.504.3168B}.

\par\indent The formation mechanism and the properties of the vertical phase spiral have been studied in great depth. \cite{Antoja2018} and \cite{Binney2018} elaborated upon the likely origin of the phase spiral. As a high-speed dwarf galaxy impacts the disk, stars are pulled away from the center of phase space and produce a high-density region in the disk with the similar phase angle in the $z-v_z$ phase space. For stars oscillating in the disk's anharmonic vertical gravitational potential, those with different vertical action $J_z$, namely, different phase space orbits, will have different oscillation frequency $\Omega_z$. This results in a differential rotation in phase space, and the perturbed stars evolve into a phase spiral feature. The $\Omega_z-J_z$ distribution depends on the vertical gravitational potential, which means that stars with similar orbital properties will produce a similar phase spiral shape. The gravitational potential will become shallower with increasing radius, resulting in different spiral shapes at different radii, and more elongated spirals in the $z$ direction. Additionally, the phase spiral winds tighter at smaller radii \citep{Bland-Hawthorn2019,2019ApJ...877L...7W,2020ApJ...905....6X,2020ApJ...890...85L}. 
Under a given gravitational potential, the phase spiral shape will evolve with time. Thus, one can infer the impact time from the shape of the phase spiral \citep{Antoja2018,Tiann2018ApJ,Li2021,Antoja2022,Frankel2022,Darragh2023}. \citet{Tiann2018ApJ} constrained the vertical perturbation to starting no later than 0.5\,Gyr ago. The aforementioned observed characteristics of the phase spiral are in good agreement with theory.

\citet{Guo2022} investigated the possible connection between the NS asymmetry of the velocity dispersion and the vertical phase spiral found by \citet{Antoja2018}.  \citet{Guo2022} established a model through the superposition of an equilibrium background and a phase spiral component to quantitatively explain the asymmetries in the number density and velocity dispersion profiles. 
\par In this work, we compile a large sample cross-matched from the recent LAMOST and {\it Gaia} data releases, to specifically analyze the features in the vertical velocity dispersion profiles ($\sigma_{v_{z}}-z$) in different bins of Galactocentric radius $R$ (or guiding center radius $R_\mathrm{g}$) and metallicity ($[\mathrm{Fe/H}]$). We investigate the variation of the asymmetric vertical velocity dispersion profiles with $R$ ($R_\mathrm{g}$) and $[\mathrm{Fe/H}]$, and finally utilize a test particle simulation to confirm the connection between the NS asymmetry of the velocity dispersion and the vertical phase spiral found in the observations.

\par\indent This paper is arranged as follows. In Section \ref{data}, we introduce the data and sample selection criteria. In Section \ref{results}, we present the results from our data analyses of the NS asymmetries, including the NS asymmetry of the vertical velocity dispersion and its metallicity dependency. We present how these results are theoretically related to the phase spiral and provide supporting observational evidence. We reproduce these results qualitatively through test particle simulation as presented in Section \ref{simu}, and detail the necessary factors in order to reproduce the characteristics similar to the observations. Finally, we provide a discussion and summarize in Sections \ref{discussion} and \ref{summary}, respectively.

\par\indent Throughout the paper, we adopt the position of the Sun as $(x= -8.34, y=0, z=0.027)$ kpc \citep{1999A&A...352..459C,2014ApJ...783..130R} with a velocity relative to the Local Standard Rest (LSR) as $(U=9.58, V=10.52, W=7.01)\,\rm km\,s^{-1}$ \citep{2015ApJ...809..145T}. 
These adopted values are well consistent with those widely recognized works \citep[e.g.,][]{2012MNRAS.427..274S,2016ARA&A..54..529B,2019AA...625L..10G}. In principle, the solar motion with respect to the LSR does not affect the velocity dispersion, the position of the Sun just slightly affects the calculation of the guiding center radius.
%{\bf Theses adopted values are consistent with the review of \cite{2016ARA&A..54..529B}, i.e., the Sun's distance from Galactic Center $R_0 = 8.2\pm0.1$\,kpc, the solar offset from local disk midplane $z_0=25\pm5$\,pc, and the solar motion with respect to the LSR $(U_\odot,V_\odot,W_\odot)=(10.0\pm 1, 11.0\pm 2, 7.0\pm 0.5)\,\rm km\,s^{-1}$. These values are most reliable as they are data summarized from multiple surveys. }

\section{Data}
\label{data}
\par\indent In order to obtain six dimensional kinematic information, we utilize the precise parallaxes and proper motions in {\it Gaia} eDR3 \citep{2021A&A...649A...1G}, and the line-of-sight velocities provided by LAMOST DR7 \citep[][]{Cui2012,Luo2015}. The cross-matched catalog between {\it Gaia} eDR3 and LAMOST DR7 contains 10,143,169 stars. We  exclude the duplicate sources by self-crossmatching and keep the star with highest signal-to-noise ratio in $g$ band ($\mathrm{SNR}_g$).

We apply the following criteria to ensure the data quality:
\begin{itemize}
    \item {\tt parallax\_err/parallax} $<0.2$,
    \item {\tt pm\_err/pm} $<0.2$,
    \item $\mathrm{SNR}_g>20$,
\end{itemize}
where {\tt pm} is the proper motion defined as $\sqrt{{\tt pm\_ra}^2+{\tt pm\_dec}^2}$.
\par We use $[\mathrm{Fe/H}]>-0.4$ as the condition to select thin disk stars, where [Fe/H] is that provided by LAMOST DR7. Note that our sample includes both main sequence and giant stars. In order to exclude halo stars, we adopt a kinematic cut in the absolute velocity relative to the local standard of rest (LRS): $|V-V_\mathrm{LSR}|<200$ km\,s$^{-1}$. The stars remaining after this step are defined as the preliminary sample (named as Sample\,0). Note that the line-of-sight velocity from LAMOST is systematically smaller than that from {\it Gaia}. We add 5.7 km\,s$^{-1}$ to our LAMOST line-of-sight velocities as a bias correction \citep{2015ApJ...809..145T}. 

\par\indent A key focus of LAMOST is to observe in the Galactic anticenter direction \citep{Deng2012,Zhao2012,Liu_Xiao-Wei_2015}, resulting in a dearth of stars at small Galactocentric radii. To ensure a sufficient number of stars for binning the data, we avoid the inner Galactic regions. We select stars located in the radial range of $8<R<11 \rm\,kpc$ and the azimuthal angle range of $-10\degr< \phi <10\degr$ in the Galactocentric cylindrical coordinates as our first sample (named as Sample\,1 throughout the paper). 
The second sample (named as Sample\,2) is defined similarly and only differs in the definition of the radius: we select stars using the guiding center radius ($R_\mathrm{g}$), which reflects the orbital properties of stars. We calculate the guiding center radius using {\tt galpy} \citep{2015ApJS..216...29B} under the Milky Way potential {\tt MWPotential2014}. The radial criterion we use to define the Sample\,2 is $8<R_\mathrm{g}<11\rm\,kpc$. Figure \ref{fig1} shows the $R-z$ and $R-\phi $ distribution of the full matched LAMOST/{\it Gaia} stellar catalog and of our selected Sample\,1 and Sample\,2. Our Sample\,1 and Sample\,2 contain 1,052,469 and 754,635 stars, respectively.
\par\indent 
The influence of the bias in distance on the phase spiral shape is discussed in \cite{Antoja2022}. Distances derived from the inverse of parallaxes suffer from an asymmetric systematic bias, which is then transferred to the calculation of coordinates and velocities. Failure to account for this effect will cause the phase spiral to shrink \citep[see the appendix of][]{Antoja2022}. To avoid this effect, we therefore adopt the distances from \cite{2021yCat.1352....0B}. 
We convert the distances, line-of-sight velocities, and proper motions to cylindrical coordinates and corresponding uncertainties by adopting the method of \cite{1987AJ.....93..864J}.
In Figure \ref{fig2} we show the $R-z$ map color-coded by the median vertical velocity error and median $z$ error.
%The uncertainty in vertical velocity is given by {\tt VrpmUVW} \citep{1987AJ.....93..864J} in the python package: astrobox. 
The median of the vertical velocity error is $\rm <5\,km\,s^{-1}$ and the median of the $z$ error is $\rm 0.06\,kpc$ in the majority of regions studied, which attests to the good quality of our data set. We note that the error increases with Galactic latitudes. This is because the line-of-sight velocity error is dominant in the vertical direction, and is larger than the error of the precise proper motions. 

\section{Results} %: 
\label{results}

\par\indent The characteristics of NS asymmetry detected in the vertical velocity dispersion ($\sigma_{v_{z}}$) profile give clues to its origin. In this section we characterise the vertical velocity dispersion profile and look for clues to a common origin with the vertical phase spiral. We elaborate upon the new features that we find within the NS asymmetry profile. Incorporating the origin of the phase spiral, we provide a qualitative explanation for the NS vertical velocity dispersion asymmetry and its detailed characteristics. Finally, we show the connection between NS asymmetries (including stellar density and metallicity) with the phase spiral.

\subsection{Peaks and troughs in $\sigma_{v_{z}}-z$ profile}
\label{concavity}

\begin{figure*}
    \centering
    %\subfigure{
    \includegraphics[width=9.0cm, trim=1.cm 1.3cm 1.6cm 2.6cm, clip]{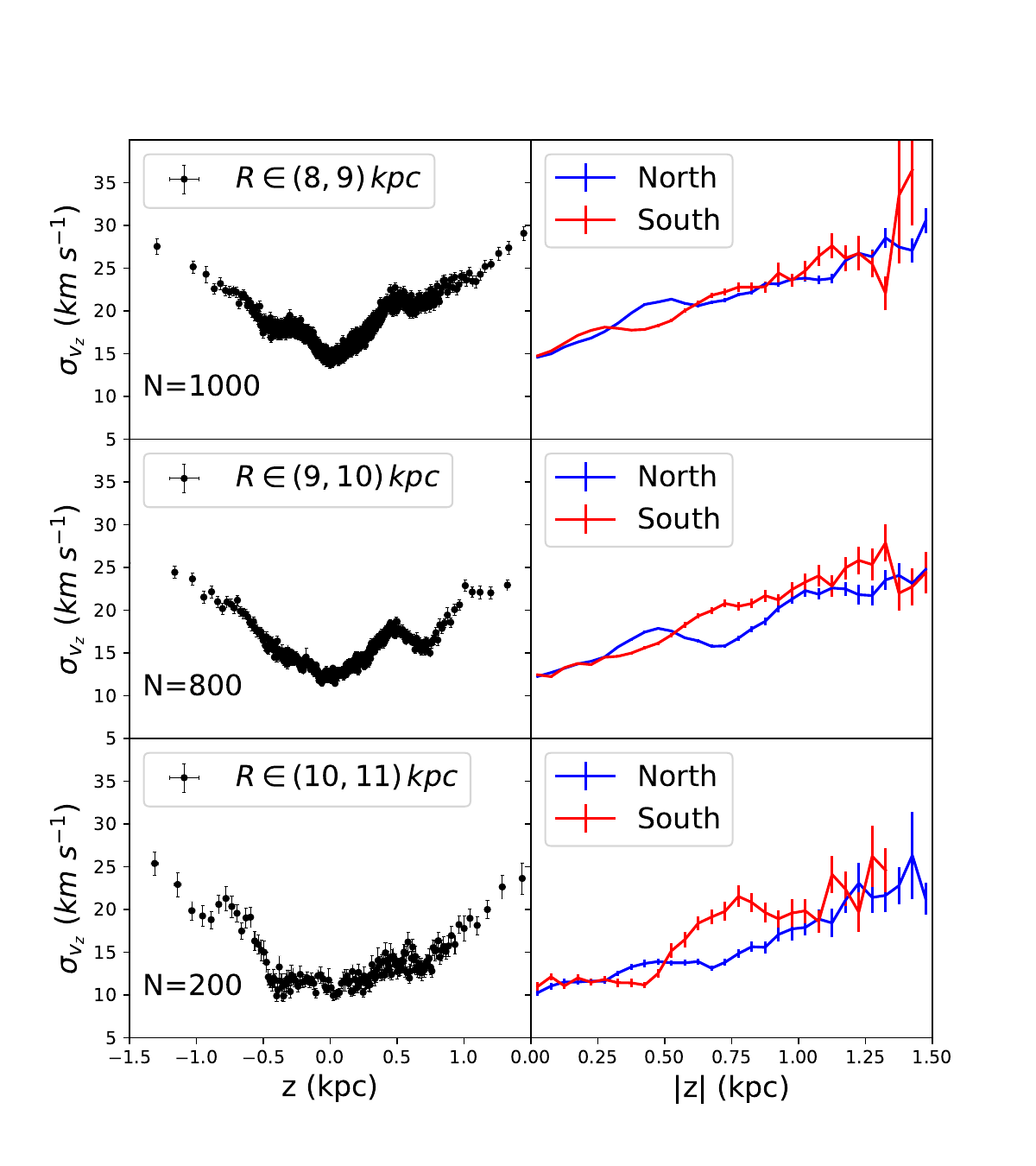}
    %}
    %\subfigure{
    \includegraphics[width=8.35cm, trim=2.3cm 1.3cm 1.5cm 2.6cm, clip]{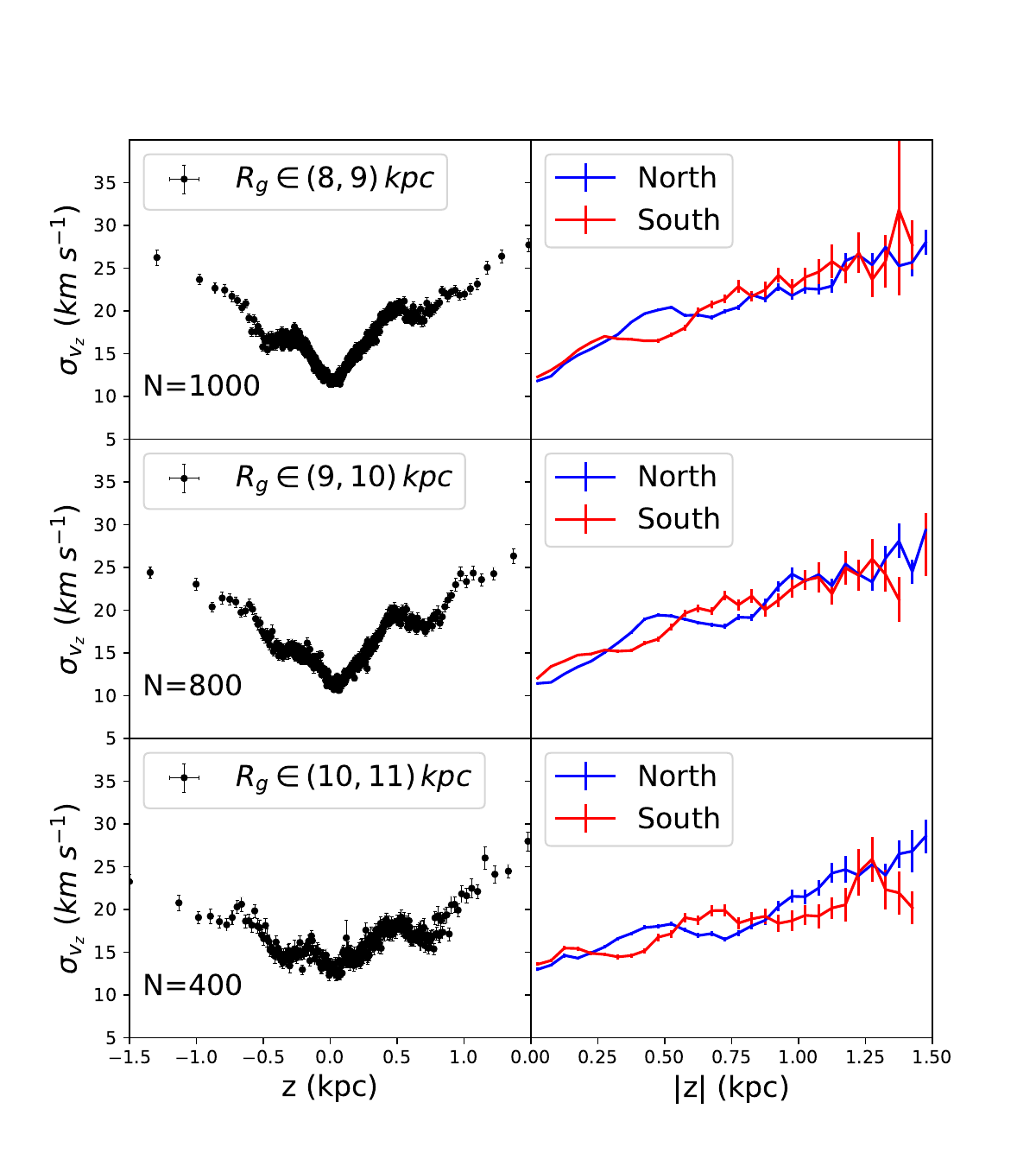}
    %}
    \caption{Left panel: Comparison of vertical velocity dispersion $\sigma_{v_z}$ profiles in different radial $R$ ranges (as indicated in the top-left corner of in each row) for Sample\,1. The left column of the panel shows the vertical velocity dispersion versus $z$. Each data point is estimated by a fixed number $N$ of stars sorted by $z$; $N$ is indicated in the bottom left of each row. The right column of the panel shows the vertical velocity dispersion profiles separately for both North (blue) and South (red) of the disk midplane binned by $|z|$. The data are binned using a linear gridding with width of 0.05\,kpc. Right panel: Same as the left panel, except using Sample\,2, thus the data are binned using the guiding center radius $R_\mathrm{g}$. Compared to the results in the left panels, the features (e.g., the troughs near $z = 0.6$\,kpc and $z = -0.5$\,kpc) are more striking in the profiles binned with $R_\mathrm{g}$ shown in the right panels}.
    \label{fig3}
\end{figure*}
\begin{figure}
    \centering
    \includegraphics[width=9cm, trim=0.5cm 1.cm 0.8cm 2.5cm, clip]{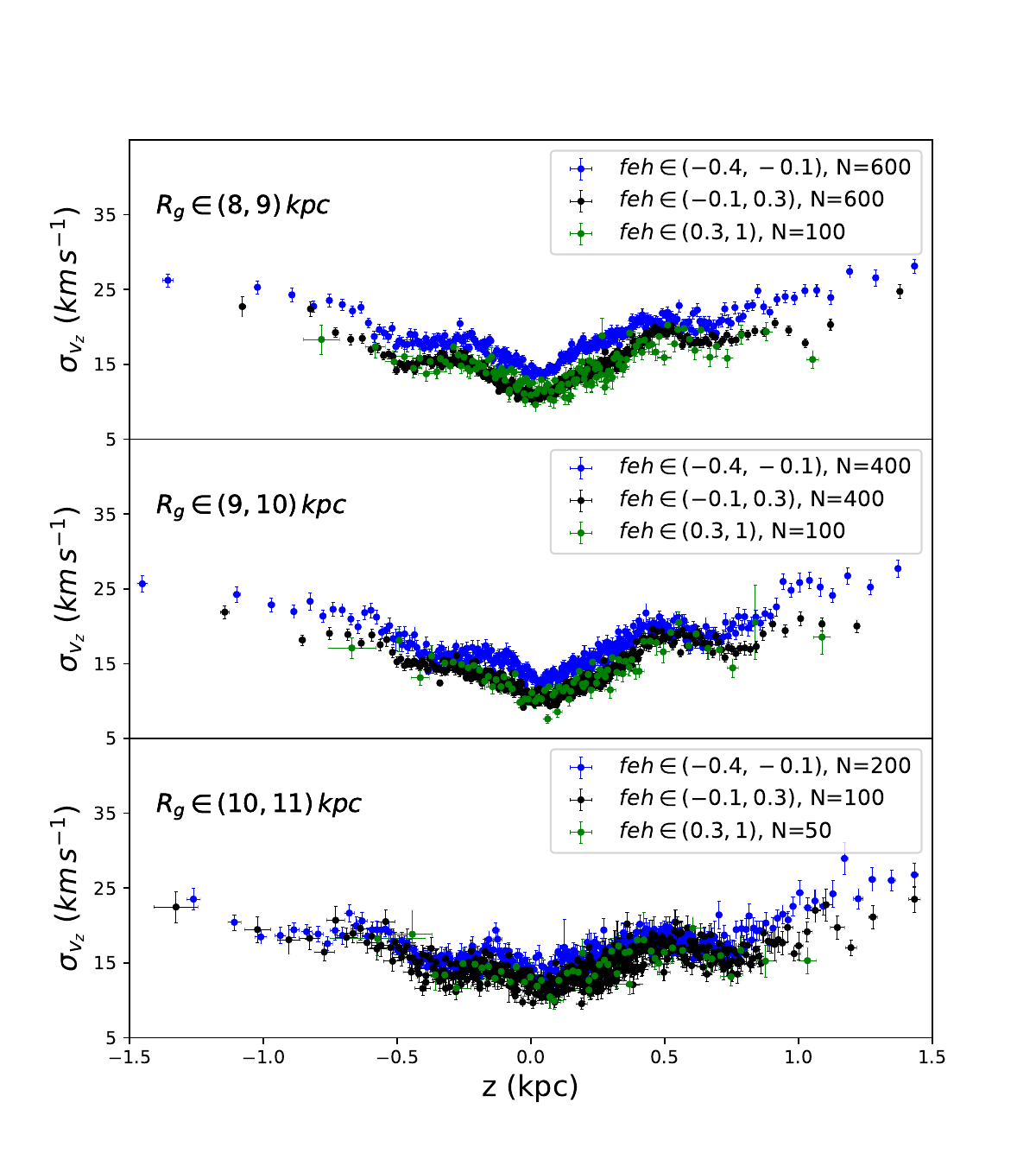}
    \caption{Vertical velocity dispersion profiles for three different guiding center radial bins of Sample\,2. The metallicity range is indicated in the legend of each panel, along with the sample size $N$ per bin. }
    \label{fig4}
\end{figure}
\par\indent 
We calculate the vertical velocity dispersion $\sigma_{v_{z}}$ from the standard deviation in $v_z$. Considering the uncertainty of vertical velocity will lead to an overestimated standard deviation, we correct the median uncertainties of $v_z$ as $\sigma_{v_{z}}=(\sigma^2 -\bar{v}_{z,\mathrm{err}}^2)^{1/2}$.
We estimate the uncertainty on $\sigma_{v_{z}}$ using the bootstrapping method.
Black data points in both the left and right panels of Figure \ref{fig3} show the vertical velocity dispersion profiles in different ranges of Galactocentric radius $R$ (the left panel) and guiding center radius $R_\mathrm{g}$ (the right panel). 
Apart from the general trend that the vertical velocity dispersion increases with increasing $|z|$, the most striking feature is the troughs of the profile near $z = 0.6$\,kpc and $z = -0.5$\,kpc.  The features are clearer in the profiles binned with $R_\mathrm{g}$. This is reasonable since the stars have similar orbits in a given $R_\mathrm{g}$ bin. The same peak and trough features are found in subsamples of different ranges in radius and metallicity, as shown in Figure \ref{fig4}. Thus, we rule out the possibility that these peaks and troughs are due to individual stellar associations. Although the shape of the $\sigma_{v_{z}}$ profile is independent of metallicity, Figure \ref{fig4} shows an overall shift in the $\sigma_{v_{z}}$ profile with metallicity.

\par\indent The phenomenological model proposed by \cite{Guo2022} is able to explain the trough features of the vertical velocity dispersion curve by associating them with the components of the asymmetric phase spiral. \citet{Guo2022} simply regards the distribution of stars in the $z-v_z$ phase space as a superposition of a smooth background and an asymmetric spiral component. 
If the spiral intersects with the $z$ axis at a certain $z$, there are more stars with $v_{z}$ close to zero velocity, which will reduce the velocity dispersion. For vertical regions away from the intersections, the velocity dispersion will increase. Therefore, the phase space spiral will cause a peak in the vertical velocity dispersion profile at the intersection between the phase spiral and the $z$ axis. In addition, stars in the process of vertical oscillation are more often at high vertical height and low velocity states, thus the phase space spiral component will be enriched near the intersection of the phase space orbit and the $z$ axis, which will enhance the trough found in the $\sigma_{v_{z}}$ profile. We expect that the peak and trough features that we find in the $\sigma_{v_{z}}-z$ curve thus trace the phase spiral. 

\begin{figure}
    \centering 
    \includegraphics[width=9cm, trim=0.0cm 1.5cm 0.8cm 1.5cm,clip]{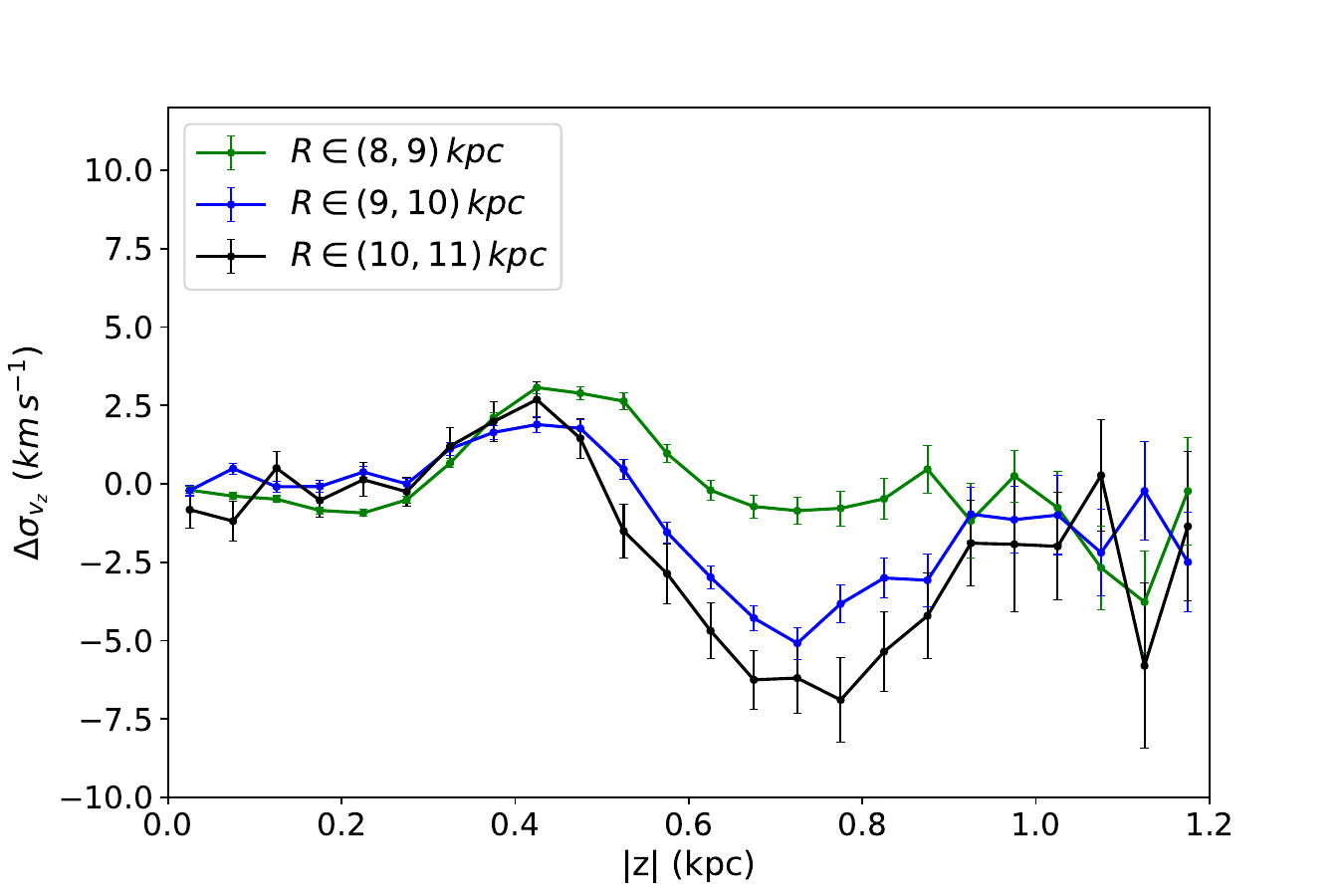}
    \includegraphics[width=9cm, trim=0.0cm 0.0cm 0.8cm 1.7cm,clip]{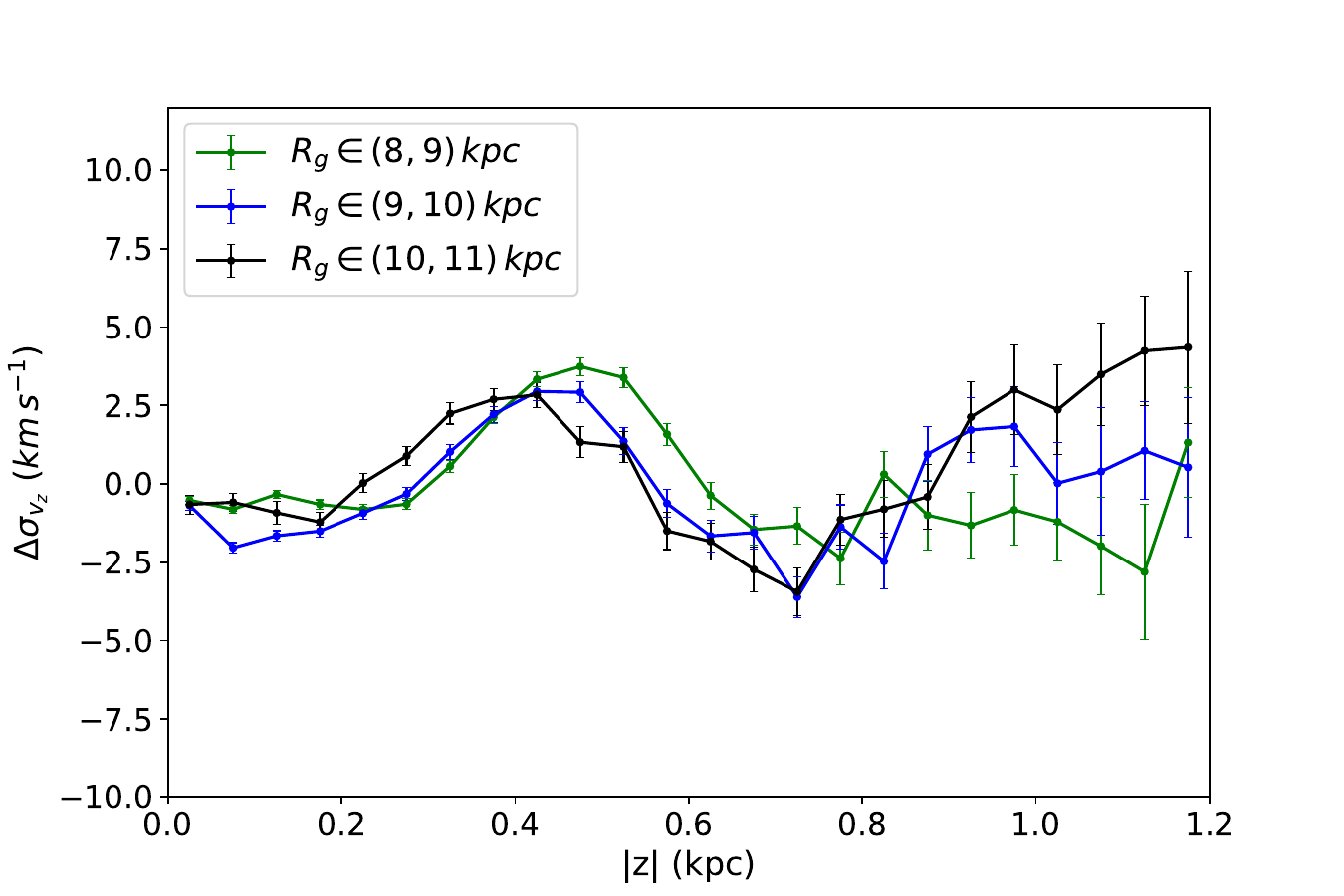}
    \caption{NS $\sigma_{v_z}$ asymmetry profiles for Sample\,1 (in different Galactocentric radial ranges, upper panel) and Sample\,2 (in different guiding center radial ranges, lower panel), where $\Delta \sigma_{v_z} = \sigma_{v_{z},\mathrm{north}}-\sigma_{v_z,\mathrm{south}}$. The green, blue, and black lines represent stars in different radial ranges as indicated in top left corner of each panel.
    }
    \label{fig5}
\end{figure}

\begin{figure*}
    \centering
    \includegraphics[width=18cm, trim=1cm 1cm 1cm 1.5cm,clip]{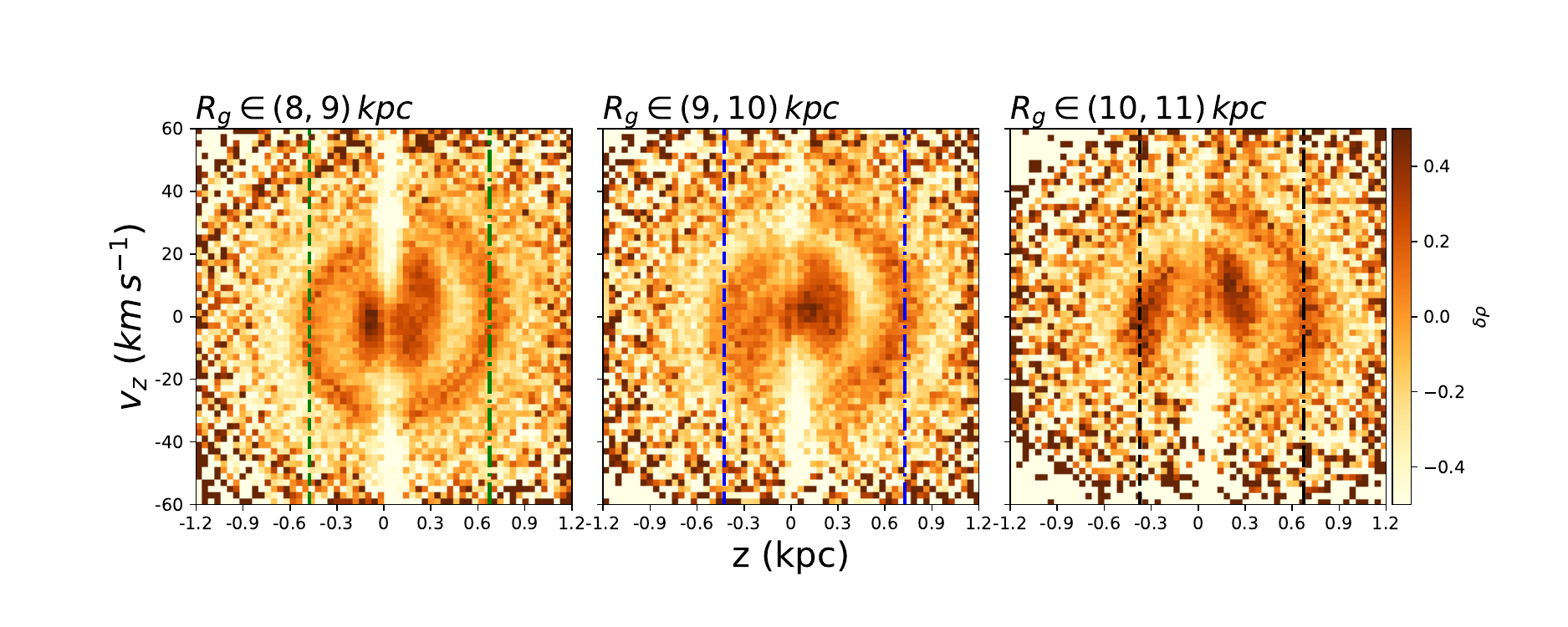}
    \caption{Phase spirals for stars with different $R_\mathrm{g}$ ranges color-coded by $\delta\rho$ for our Sample\,2. The dashed lines indicate the intersections between the phase spirals and $z$ axis.}
    \label{spiral_Rg}
\end{figure*}

\begin{figure*}
    \centering
    \includegraphics[width=18cm, trim=0.0cm 1.5cm 1cm 1.5cm,clip]{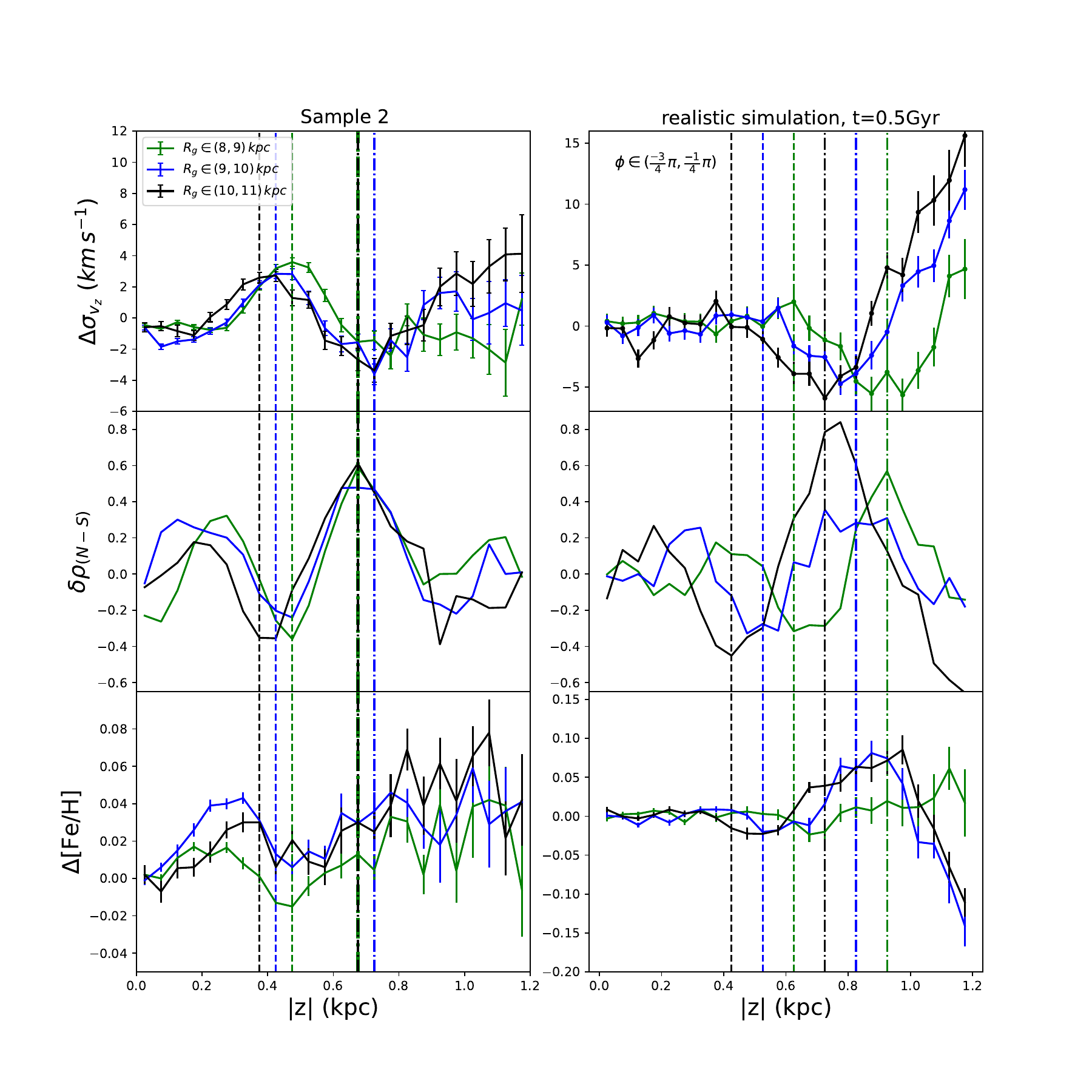}
    \caption{NS difference profiles from our Sample\,2 (left column) compared to our test particle simulation (right column). The solid lines show profiles with different $R_\mathrm{g}$ ranges, as indicated in the legend. The vertical dashed lines indicate the intersections between the phase spirals with different $R_\mathrm{g}$ and the $z$ axis, which are the same for Sample\,2 as indicated in Figure \ref{spiral_Rg}. First row: NS difference profiles of vertical velocity dispersion, i.e. $\Delta \sigma_{v_z}$. Second row: NS difference profiles of number density ($\delta\rho$), for stars within $|v_z|<10\,\rm km\,s^{-1}$. Third row: NS difference profiles of median [Fe/H]. The bins along $|z|$ are 0.05\,kpc. For the realistic simulation on the right, we allow the sample to evolve for 0.5\,Gyr, with an azimuthal range of $[\frac{-3}{4}\pi,\frac{-1}{4}\pi]$.
    }
    \label{dns_rho_feh}
\end{figure*}

\begin{table*}
 
    \centering
    \caption{Pearson's correlation between NS asymmetry curves in different radial ranges. The uncertainties are estimated by the technique of bootstrapping.}
    \renewcommand\arraystretch{2}
    \begin{tabular}{lclclcl}
    \toprule
            &\multicolumn{2}{c}{$8<R_\mathrm{g}\mathrm{/kpc}<9$}&\multicolumn{2}{c}{$9<R_\mathrm{g}\mathrm{/kpc}<10$} &\multicolumn{2}{c}{$10<R_\mathrm{g}\mathrm{/kpc}<11$}\\
    \hline
     &$P_\mathrm{corr\_coeff}$ &$p$-value &$P_\mathrm{corr\_coeff}$ & $p$-value &$P_\mathrm{corr\_coeff}$& $p$-value\\
    \midrule
    
           $\delta\rho\, vs.\, \Delta \sigma_{v_z}$ & $-0.647_{-0.053}^{+0.030}$ & $2.04_{-1.45}^{+1.72}\times 10^{-3}$ & $-0.856_{-0.035}^{+0.028}$ & $1.48_{-1.35}^{+5.24}\times 10^{-6}$ & $-0.759_{-0.081}^{+0.050}$ & $1.05_{-1.01}^{+3.60}\times 10^{-4}$\\
    \hline
           $\delta\rho\, vs.\, \Delta [\mathrm{Fe/H}]$ & $0.595_{-0.029}^{+0.065}$ & $5.60_{-4.10}^{+3.78}\times 10^{-3}$ & $0.594_{-0.048}^{+0.050}$ & $5.73_{-3.54}^{+6.73}\times 10^{-3}$ & $0.231_{-0.079}^{+0.062}$ & $1.83_{-0.77}^{+1.42}\times 10^{-1}$\\
    \bottomrule
    \end{tabular}
 \begin{tablenotes}
\item Note: The error bars and $p$-values we calculate using a bootstrapping method that does not account for the significant autocorrelation of the signals, implying that both error bars and $p$-values are underestimated.
%\item $^1$ Note: $R_\mathrm{g}$ is in the unit of kpc.
 \end{tablenotes}
    \label{tab1}
\end{table*}

\subsection{NS asymmetry of $\sigma_{v_{z}}$}
\label{NSa_svz}
\par\indent Since the heights of the intersections between the phase spiral and the $z$ axis are different in the North and South, the heights of the peaks and troughs in the $\sigma_{v_{z}}$ profile are different between the North and South, as shown by the blue and red curves in the left ($R$ bins) and right ($R_\mathrm{g}$ bins) panels of Figure \ref{fig3}. This results in the NS asymmetry of vertical velocity dispersion. We now investigate the difference between the North and South velocity dispersion profiles by comparing our samples in bins of different $R$ and $R_\mathrm{g}$, and in Section \ref{metal}, bins of [Fe/H]. We find that such comparisons further highlight the characteristics of the NS asymmetry.

\par\indent Figure \ref{fig5} shows the difference between the North and South velocity dispersion profiles, i.e., 
$\Delta \sigma_{v_{z}}=\sigma_{v_{z},\mathrm{north}}-\sigma_{v_{z},\mathrm{south}}$
(we refer to this as the NS asymmetry curve of the velocity dispersion profile), binned along $R$ (upper panel) and $R_\mathrm{g}$ (lower panel). The NS asymmetry curve is characterized by a wave feature which first shows a peak then trough with increasing $|z|$. We find that the same wave feature of the NS asymmetry curve appears in all three radial bins investigated, although the feature is slightly shifted along $|z|$. The wave at smaller $R$ ($R_\mathrm{g}$) curves is shifted to increasingly larger values of $|z|$. This shift is clearer in the asymmetry curves binned with $R_\mathrm{g}$. The intensity of the NS asymmetry produced by the phase spiral component is affected by the ratio of the stellar number in the spiral components and the background. The NS asymmetry of velocity dispersion near the midplane of the disk is not obvious for different $R$ ($R_\mathrm{g}$) bins probably because the ratio of the spiral component to background is small.

\par\indent If we assume that the asymmetry of the vertical velocity dispersion is caused by the phase spiral, and combine this with the formation mechanism and characteristics of the phase spiral, we are able to qualitatively explain the shifts seen in the curves at different $R$ ($R_\mathrm{g}$) in Figure \ref{fig5}.
The formation of the phase spiral originates from the anharmonicity of the vertical oscillation of stars in the disk \citep{Antoja2018}. The oscillation frequency of stars decreases with the increase of $J_z$. In other words, in the $\rm z-v_z$ phase space, the angular velocity of a star near the origin is faster, while the angular velocity of a star far from the center is slower. The spiral feature comes from the differential rotation of stars with different $J_z$. Since the $J_z-\Omega_z$ depends on the vertical gravitational potential profile, the evolution of the phase space spiral fundamentally depends on the vertical gravitational potential experienced by these stars. Therefore the phase spiral shapes at different Galactic radii are different due to the different vertical potential \citep{2018MNRAS.477.2858W}. However, stars at the same $R$ may have different orbital properties, such that the guiding center radius is a better way to distinguish phase spirals with different intrinsic shapes \citep{2020ApJ...890...85L,2021MNRAS.508.1459H}. Separating the samples according to $R_\mathrm{g}$ to get phase spiral components with the same intrinsic shape reduces the morphological ambiguity brought about by mixing. Thus, the shift in the NS asymmetry curves are clearer when samples are binned by $R_\mathrm{g}$. Stars with smaller $R_\mathrm{g}$ experience steeper vertical gravitational potential, so their oscillation frequency is generally higher and they tend to rotate with larger phase angles at the same time interval. Therefore, the spiral consists of stars with small $R_\mathrm{g}$ located in the outer edge, which is well shown in Figure 18 of \citet{Li2021}, so that the vertical location of the $z$-axis intersection is larger. The NS asymmetric curve of stars with smaller $R_\mathrm{g}$ will be biased to a relatively larger $|z|$.

\subsection{NS asymmetries and their connection with the phase spiral}
\label{spiral_sigmaz_feh}
In the previous sections, we have detailed the NS asymmetry of the vertical velocity dispersion observed in our sample and have made inferences under the assumption of a direct connection to the vertical phase space spiral. Here we explore the NS asymmetries in number density and [Fe/H] and their connection with the phase spiral. Our results are summarized in Figure \ref{spiral_Rg} and the left column of Figure \ref{dns_rho_feh}.
\par We define the density contrast as $\delta\rho= n/n_{G}-1$, where $n$ is number count of stars in the phase space cells in ($z,v_{z}$) and $n_{G}$ is the count smoothed by a Gaussian filter with $\sigma_{G}=3$ cell width. Figure \ref{spiral_Rg} shows the phase spirals within three $R_\mathrm{g}$ bins color-coded by $\delta\rho$. The dashed lines show the locations of the intersection between the spiral and $z$ axis (same lines plot in the left panels of Figure \ref{dns_rho_feh}). The first row of Figure \ref{dns_rho_feh} shows the $\Delta \sigma_{v_{z}}$ profile in three $R_\mathrm{g}$ bins, which is same as the bottom of the Figure \ref{fig5}. In the second row of Figure \ref{dns_rho_feh}, we use $\delta\rho_{(N-S)}$ to represent the NS asymmetry in number density, which is the difference of $\delta\rho$ between the North and South for a stripe in phase space with $ |v_{z}|<10\,\rm km\,s^{-1}$. The lower panel left column of Figure \ref{dns_rho_feh} shows the NS asymmetry curve for [Fe/H]. The peaks in the NS asymmetry of the vertical velocity dispersion roughly correspond to the trough in the NS asymmetry of number density and metallicity. As indicated by the dashed lines, the common shift shown in the NS asymmetry profiles of both $\sigma_{v_{z}}$ and $\delta\rho$ is strong evidence of a common origin due to the phase spiral.

%\textbf{We utilize the similarity between the NS asymmetry curves to show their correlation. After normalizing the difference between the maximum and minimum values to unit one, we estimate the similarities between the NS asymmetry curves using their Fr\'echet distance \citep{2012arXiv1204.5333A}. We note that the velocity dispersion curve is inversed compared to the shape of the density and metallicity curves. We calculate the similarity between the NS asymmetry of density and velocity dispersion, and the similarity between the NS asymmetry of number density and metallicity. We exclude the data in the range of $|z|>1.0$\,kpc due to the large uncertainties. As shown in Table \ref{tab1}, the measured similarity values are high in all $R_\mathrm{g}$ bins. This is strong evidence of the correlation between the NS asymmetry in velocity dispersion, metallicity and number density and their shared origin stemming from the phase spiral.}

\par Similar to the vertical velocity dispersion distribution and density distribution, the vertical distribution of metallicity abundance also shows a NS asymmetry (lower panel of Figure \ref{dns_rho_feh}). The peaks and valleys of the NS asymmetry wave pattern of $\sigma_{v_{z}}$ are at similar $|z|$ oppositely fluctuating for $\delta\rho$ and [Fe/H]. The oppositely fluctuating asymmetry in metallicity were first reported by \cite{2019ApJ...878L..31A}. At the similar location where the wave feature in the vertical velocity dispersion NS asymmetry curve shows a peak, there is a valley in the NS difference of both density and [Fe/H]. This phenomenon is a natural consequence of the model describing the asymmetry as a product of the vertical phase spirals excited by disturbances. The phase spiral component is composed of stars that have been disturbed by the gravitational potential of a perturber. Stars in the central part of the $z-v_z$ phase space have higher metal abundance. After obtaining a perturbation in velocity, the stars gain energy and enter a phase space orbit with higher $J_{z}$. As the orbits of these excited stars evolve, these stars become the main components of the phase spiral and they have higher [Fe/H] than the undisturbed background components of the disk with the same $J_{z}$. Here we ignore the effect caused by the radial disturbance and only consider the change of the metal abundance distribution caused by the vertical velocity excitation. As mentioned earlier in Section \ref{concavity}, vertically oscillating stars stay longer at low velocity states, thus the number of spiral components is higher near the $z$ axis in vertical phase space. As a result, the increase of median metal abundance is more obvious.
\par The spiral also can be displayed in the $z-V_z$ phase space by color-coding based on metal abundance. \cite{2022arXiv220605534G} used [M/H] residuals to color-code the vertical phase space distribution. They also suggest that this NS asymmetry feature is possibly connected with the phase spiral.
\par If the wave signal in the NS asymmetry of number density originates from the phase spiral as we assume in this study, the correlation between the NS asymmetry of number density and other asymmetries will reflect how much the phase spiral is able to account for this NS asymmetry signal. To inspect this correlation, we calculate the Pearson correlation between the NS asymmetry of density and velocity dispersion, and the Pearson correlation between the NS asymmetry of number density and metallicity. As this statistic measures the linear correlation coefficient, such correlation does not precisely represent the relation between the NS asymmetry profiles, but is enough to allow us to make a qualitative judgement.
\par Table\,\ref{tab1} shows the correlation between the NS asymmetry curves in different ranges of guiding center radius. In our calculations of the correlations, we exclude the data with $|z|>1.0$\,kpc due to the large uncertainties. We firstly build 50 subsamples, each one consisting of a random selection of $60\%$ of the stars from our full sample. We calculate 50 sets of NS asymmetry profiles from different $R_\mathrm{g}$ bins using the same process as in previous sections. Then we calculate the Pearson correlation between the NS asymmetry of number density and metallicity from each set of subsamples. Finally we adopt the 50th percentile as our result and 16th and 84th percentiles as our error bar. The error bars and $p$-values we calculate using this bootstrapping method does not account for the significant autocorrelation of the signals; this implies that both error bars and $p$-values are underestimated.

\par The high correlation between the NS asymmetry of number density and vertical velocity dispersion shows their strong connection. The strong connection is also supported by the close agreement of their peak/trough locations, as seen by the dashed lines of Figure \ref{dns_rho_feh}. The correlations between the NS asymmetry of number density and metallicity are low in $R_\mathrm{g}\in (9,11)$\,kpc, while we find a moderate correlation in $R_\mathrm{g} \in (8,9)$\,kpc. On the one hand, the wake-like signal may be smoothed out by the uncertainties on [Fe/H]. This could weaken the peaks and troughs at large radial range. On the other hand, there is a trend that the NS difference varies with $|z|$, which can decrease the {\it Pearson} correlation coefficient ($P_\mathrm{corr\_coeff}$). Our analysis with Pearson's correlation statistic supports our assumption that the wave-like signal is caused by the phase spiral, but is only detected near the solar radius. 

\subsection{Metallicity dependence of NS asymmetry of $\sigma_{v_{z}}$ and phase spiral}
\label{metal}

\begin{figure*}
    \centering
    %\subfigure{
    \includegraphics[height=4.5cm,width=18cm, trim=1.5cm 1.4cm 1.5cm 1.5cm, clip]{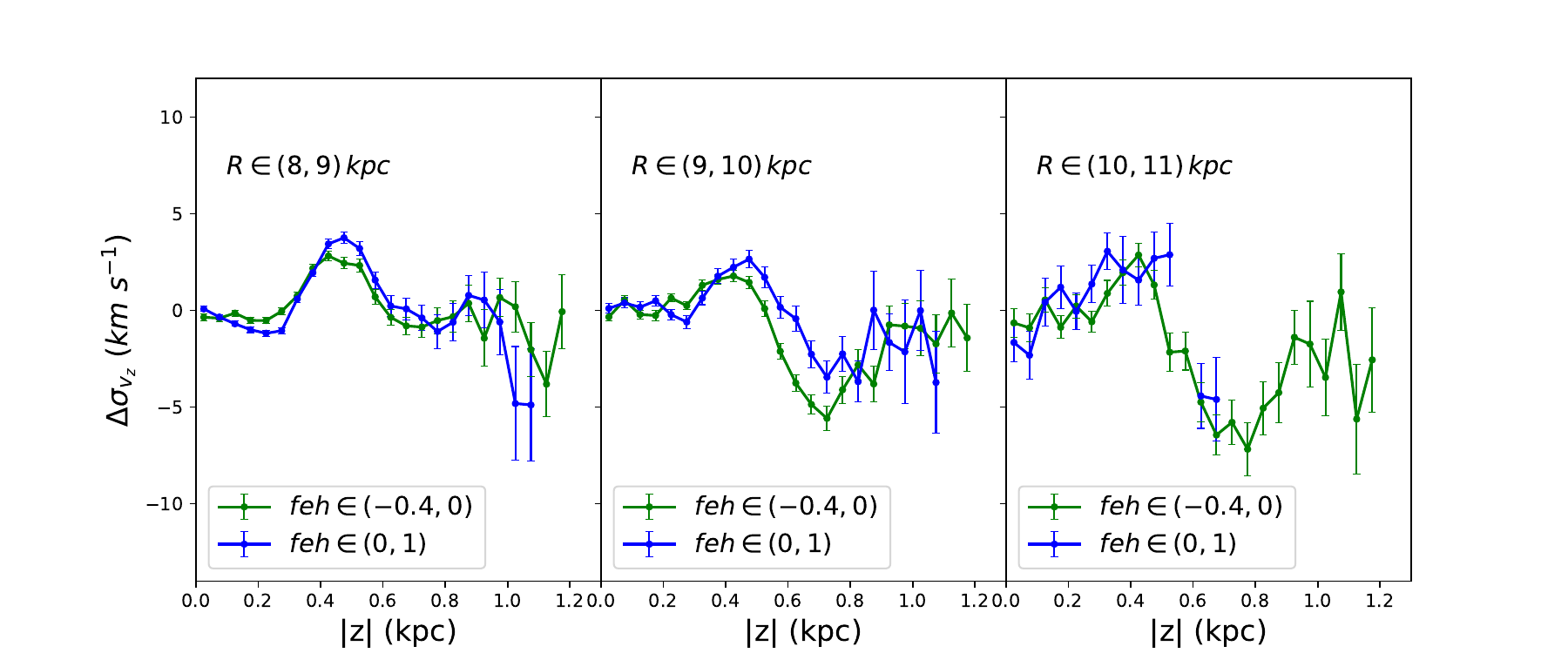}
   % }
    %\subfigure{
    \includegraphics[height=5.0cm,width=18cm, trim=1.5cm 0.0cm 1.5cm 1.4cm, clip]{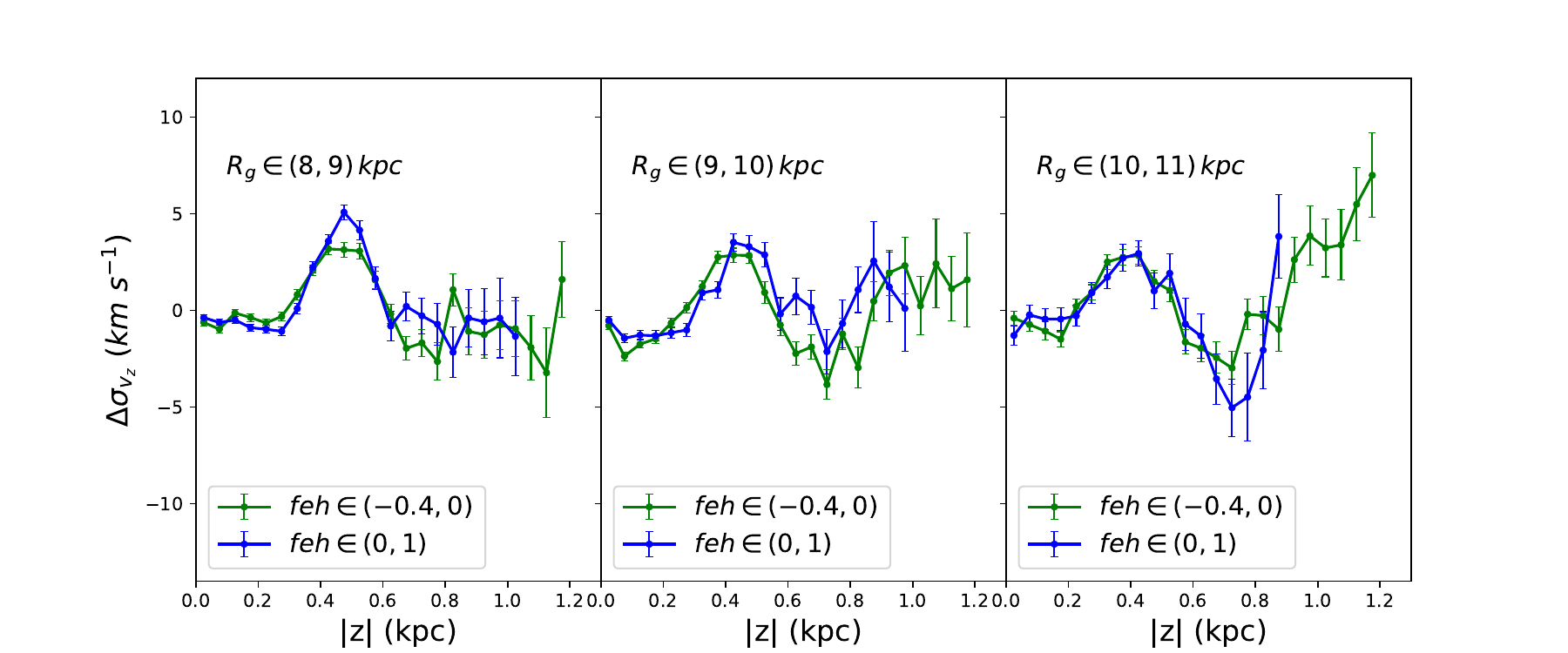}
    %}
    \caption{Observed NS asymmetry of the vertical velocity dispersion profiles along different radial ranges (Left, middle, and right columns) for Sample\,1  (Galactocentric radius, top) and Sample\,2 (guiding center radius, bottom) separated also into different bins of [Fe/H]. Green and blue curves represent stars with metallicity $-0.4<[\mathrm{Fe/H}]<0$ and $0<[\mathrm{Fe/H}]<1$, respectively.}
    \label{fig6}
\end{figure*}

Figure \ref{fig6} shows the NS asymmetry curves of vertical velocity dispersion for Sample\,1 (upper row, binned by $R$) and Sample\,2 (lower row, binned by $R_\mathrm{g}$) separated into two [Fe/H] bins. In the top panels of Figure \ref{fig6}, the wave feature tends to shift to higher $|z|$ with increasing [Fe/H], while the bottom panels slightly show such a tendency in the range of $9< R_\mathrm{g} < 10$ kpc (this only slight shift in the $R_\mathrm{g}$-binned samples is due to the stars sharing more similar orbital properties than those binned by $R$). For stars with different metallicities, their $R$ ($R_\mathrm{g}$) are generally different. The orbital properties of the stars affect the shape of the spiral, and thus lead to the differences in the NS asymmetric curves. For stars with higher [Fe/H], $R_\mathrm{g}$ is generally smaller due to the negative radial metallicity gradient. The vertical potential they experience is steeper, resulting in a phase spiral located at the outer region of phase space. Therefore, the wave pattern in the NS asymmetry curves of the velocity dispersion shifts to larger $|z|$. Nevertheless, as the samples are constrained to the same $R$ or $R_{g}$ bins, the shift of peaks between different metallicity bins is not as clear as those of the different radial bins shown in Figure \ref{fig5}.

\begin{figure*}
    \centering
    \includegraphics[width=12cm,trim=0.5cm 1cm 1cm 1.5cm, clip]{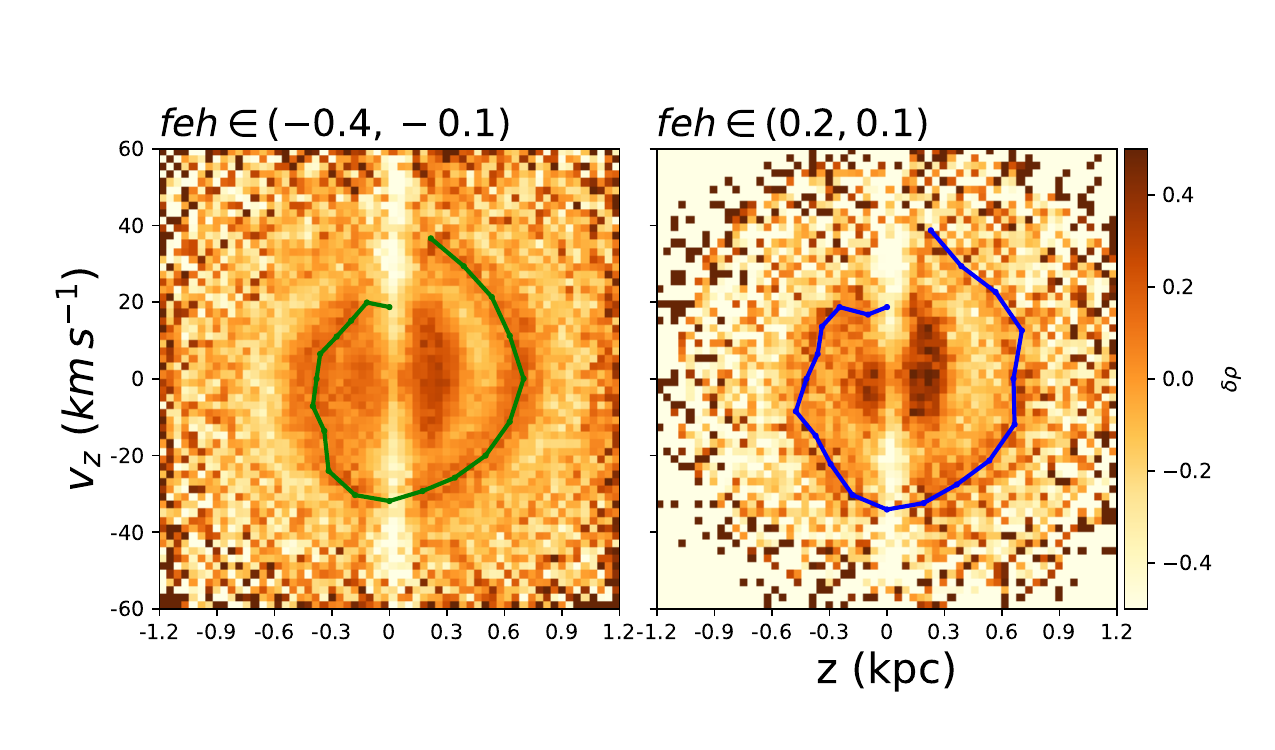}\includegraphics[width=6.0cm, trim=0cm 0cm 0cm 0.6cm, clip]{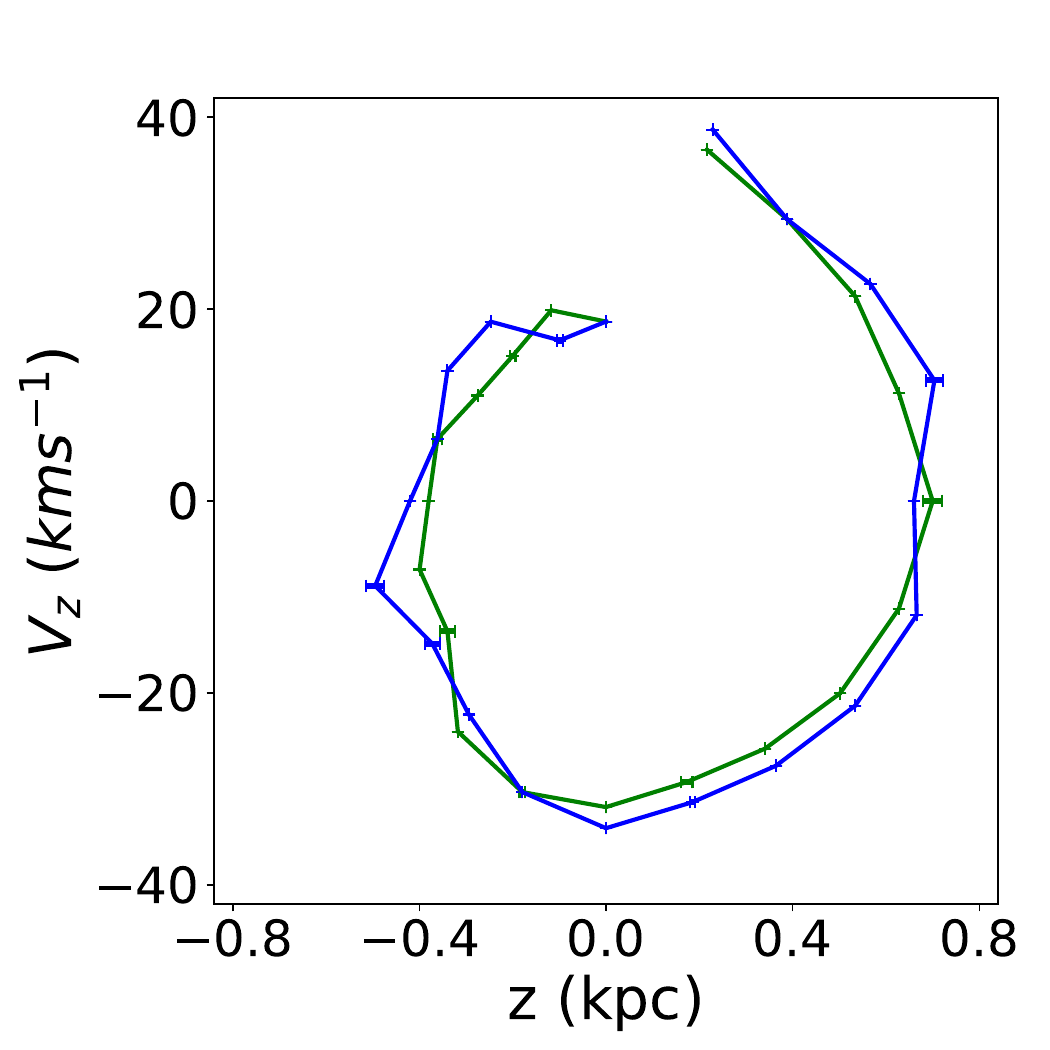}
    \caption{Observed vertical phase spiral density maps and their shape curves (green and blue lines) in different metallicity bins. Left and middle panels: Phase spirals of stars with different [Fe/H] color-coded by $\delta\rho$ for our Sample\,0 stars (only satisfying our quality criteria). The [Fe/H] range is indicated above the panels. The width of the phase grid is $\Delta z=0.04$\,kpc, $\Delta v_z=2\,\mathrm{km\,s}^{-1}$. To obtain the spiral shape curve, the phase spiral is divided into 20 sectors and the peaks of each sector are connected to form a curve (green or blue according to the given metallicity range).
    We find the peaks of the spiral in the $\theta$ range of $[\frac{1}{2}\pi,\frac{5}{2}\pi]$. We exclude the central region because the shape of the spiral is blurred. Right panel: Peaks of 20 sectors for spirals with different [Fe/H], same as the colored lines in left two panels, overplotted in one panel for comparison.}
    \label{spira_fs}
\end{figure*}

Using the preliminary sample (i.e., Sample\,0 defined in Section \ref{data}) with an additional criterion in azimuth ($-10\degr<\phi<10\degr$), we investigate how the shape of the phase spiral changes with metallicity. Figure \ref{spira_fs} shows the spirals within two metallicity bins. To obtain the shape of the phase spiral, we evenly divide each spiral into 20 sectors, and depict the shape of phase spiral by the density peaks of each of the 20 sectors. We normalize the polar coordinates of the phase space by $r=\sqrt{(v_z/55\ \mathrm{km\,s}^{-1})^2+(z/1\ \mathrm{kpc})^2}$. The phase space is divided into a grid of $(\theta, r)$ with an $r$ width of 0.04\,kpc. The $(\theta, r)$ with the maximum mean $\delta\rho$ in each sector is adopted as the coordinates of the density peak. We randomly resample 50 times with a ratio of 70\% to calculate the median values and the uncertainties of $r$. Finally, we convert the uncertainties to the errors in $z,v_z$.

As shown in the right panel of Figure \ref{spira_fs}, the spiral consisting of metal-rich stars (blue curve) is generally located on the outward side of the metal-poor spiral (green curve). The metal-rich stars tend to dominate the outer edges along the phase spiral curve. This is a reflection of the radial metallicity gradient, as the metallicity bin can be regarded as a bin in $R_\mathrm{g}$ ($V_{\phi}$). Stars with poorer metallicity are then related to a shallower vertical potential. This is consistent with the result in \cite{Antoja2022} that the shape of the phase spiral varies with $R_\mathrm{g}$.
%We focus on the spiral shapes of samples binned by metallicity as \cite{Antoja2022}, and the results are qualitatively consistent with previous works, e.g. \cite{Bland-Hawthorn2019}.} 
These results are also consistent with the shift in the wave pattern of the NS asymmetry $\sigma_{v_{z}}$ curves, because the peaks and troughs of the $\sigma_{v_{z}}$ profile are actually the intersections between the phase spiral and $z$ axis.
%\textbf{From this perspective, our results are consistent with those of \cite{Antoja2022} and reinforce the inference of \cite{} concerning the shape of the phase spiral.}


\section{Test Particle Simulations}
\label{simu}
\par\indent In this section, we use test particles to simulate the evolution of a disturbed thin disk in the gravitational potential of the Milky Way, we aim to verify whether the phase spiral generated by external disturbances could produce the observed NS asymmetry in velocity dispersion and [Fe/H]. We note that simple test particle simulations cannot reproduce the self gravity effect such as in $N$-body simulations. Though the results are not quantitative fits to the observations, they are sufficient enough to produce qualitatively comparable results. For our initial simple simulations, using test particles is much more time and cost efficient compared to a full $N$-body simulation. With these caveats, we proceed to employ the use of test particle simulation. 
\subsection{Models adopted in the simulations}
%\par Following \cite{Li2021}, we use two methods to introduce a vertical velocity perturbation to the disk test particles. In the first method, we uniformly add a vertical velocity kick to all the particles. This vertical velocity perturbation obeys a Gaussian distribution $V_\mathrm{kick}=\mathcal{N}(\mu_{v_z,\mathrm{kick}},\sigma_{v_z,\mathrm{kick}})$ where $\mu_{v_z,\mathrm{kick}}$ and $\sigma_{v_z,\mathrm{kick}}$ are the mean and dispersion in km\,s$^{-1}$. We test several strengths of perturbations and compare the results further in this section. We use this model to study the NS asymmetries caused by a simple vertical velocity disturbance and to analyze subsequent evolution of the asymmetries under the Galactic potential. 
\par\indent We mimic the impact of an intruder by adopting the perturbation model under the impulse approximation given by \cite{Binney2018}. Our goal of using this simulation is to reproduce the observed variation with $R$ in the NS asymmetry of $\sigma_{v_{z}}$. The mass and velocity of the intruder are $2\times10^{10}$ M$_\odot$ and 300\, km\,s$^{-1}$, respectively. The impact point is at $(x,y)=(20,0)$\,kpc and the impact angle is $90\degr$. The passage time scale is $T=66$\,Myr. The in-plane disturbance velocity $\delta v_{||}$ is simplified as the product of the disturbance time scale $T$ and the relative acceleration, which is reduced by the acceleration at the Galactic centre. The vertical disturbance velocity is calculated as $\delta v_{\perp}= \alpha\ \delta v_{||}[1-\beta \sin(\theta-\theta_\mathrm{Intruder})]$, where we set $\alpha=1$ and $\beta=1$. 
\par In our simulations, we adopt model I of \citet{Irrgang2013} for the gravitational potential of the Milky Way and use {\tt AGAMA} to generate thin disk particles that obey a quasi-isothermal distribution function \citep{2010MNRAS.401.2318B,2011MNRAS.413.1889B}. The radial exponential disk scale length is set as 3.7\,kpc, and the vertical scale height is set to 0.3\,kpc \citep{2015MNRAS.454.3653B,2016ARA&A..54..529B}. We generate approximately 1.5 million test particles in the range of $5<R<15$ kpc and implement orbital integration with {\tt galpy}.  

\begin{figure*}
    \centering
    \includegraphics[width=18cm, trim=0.5cm 0.5cm 1.5cm 1.0cm, clip]{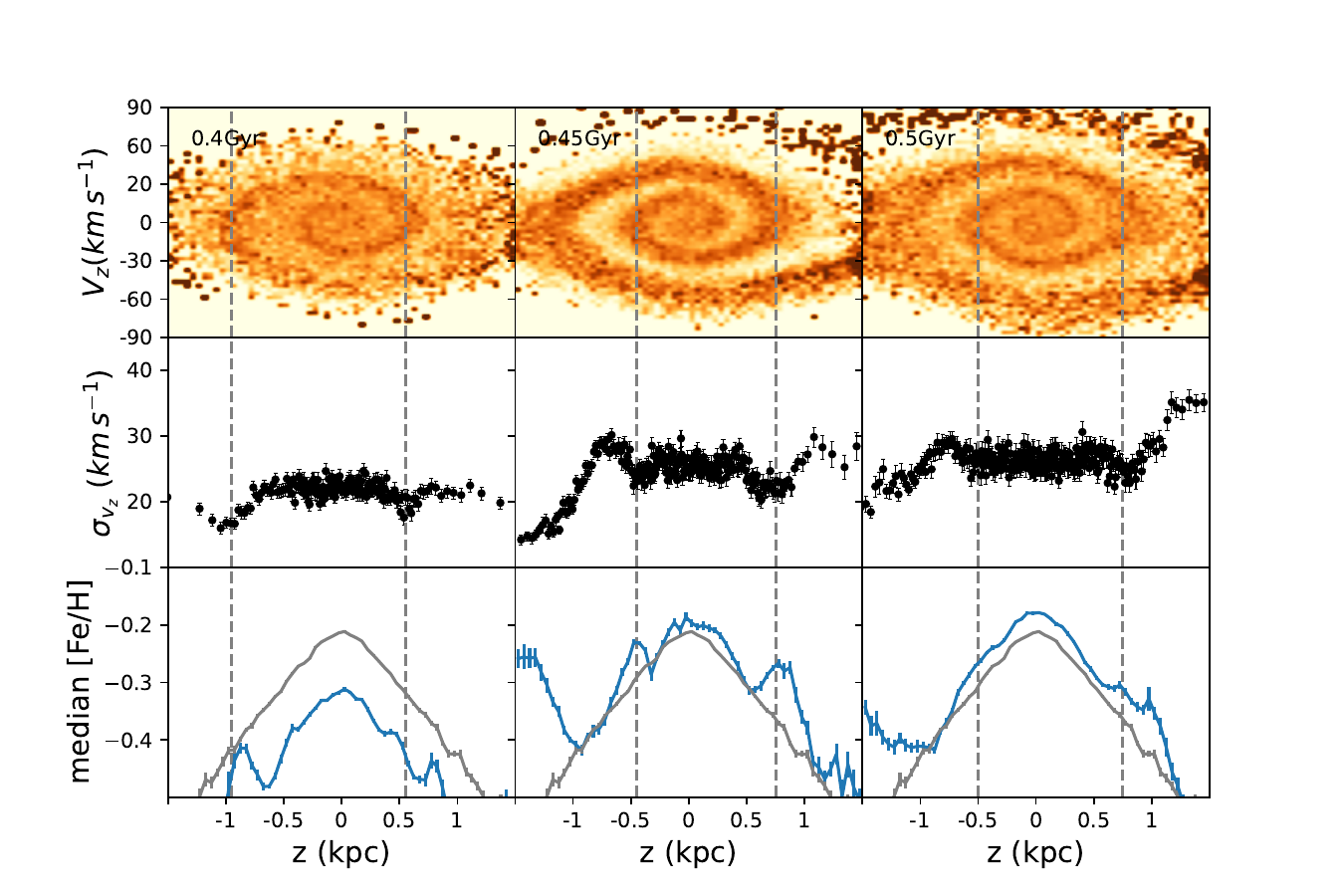}
    \caption{Results from the test particle simulation. Top row: number density distribution in phase space. Middle row: vertical velocity dispersion profile with error bars. Bottom row: median [Fe/H] profile. The grey lines represent the stable [Fe/H] profiles before we add the vertical velocity kick. The blue lines correspond to the median [Fe/H] profiles after implementing the velocity kick. The first, second, and third columns correspond to the evolution times at 0.4\,Gyr, 0.45\,Gyr, and 0.5\,Gyr, respectively. The vertical dashed lines highlight the coincidence between the phase spiral, the peaks and troughs in the vertical velocity dispersion profiles and the bumps in the [Fe/H] profiles. The radial range of the sample is $9<R_\mathrm{g}<10$ kpc and the azimuthal range is $\phi\in(-\frac{3\pi}{4} ,-\frac{\pi}{4})$.}
    \label{simu_3s}
\end{figure*}

%\begin{figure}
%\centering
%    \includegraphics[width=8cm,trim=0.6cm 0.2cm 1.5cm 1.5cm, clip]{figs/3dvz_compare.pdf}
%    \caption{NS asymmetry of vertical velocity dispersion profiles from the test particle simulations implementing the first method. The sample's evolution time is 0.5\,Gyr, and the radial range is $9<R<10$ kpc. Green, blue, and black curves represent the vertical disturbance velocity of $- 25\, \rm km\ s ^ {- 1}$, $- 30\, \rm km\ s ^ {- 1}$ and $- 35\,\rm km\ s ^ {- 1}$, respectively.}
%    \label{dvz_compare}
%\end{figure}

%\begin{figure*}
%    \centering
%    \includegraphics[width=18cm, trim=2.0cm 0.8cm 1.5cm 1.5cm, clip]{figs/simu_2model_0p5Gyr_shift.pdf}
%    \caption{NS asymmetry profiles of vertical velocity dispersion from two simulations at an evolution time of 0.5\,Gyr. Left panels show the results from our first method of vertical velocity kick and right panels show the results from our second method with the more realistic velocity kick. The results from our second method are shown for azimuth $\phi\in(\frac{-3}{4}\,\pi ,\frac{-1}{4}\,\pi)$. The red, green, and blue curves in the upper panels differ by galactocentric radial range and in the lower panels differ by guiding center radial range as indicated in the legend.}
%    \label{2_model}
%\end{figure*}

\begin{figure*}
    \centering
    \includegraphics[width=18cm, trim=1.8cm 1cm 1.2cm 1cm, clip]{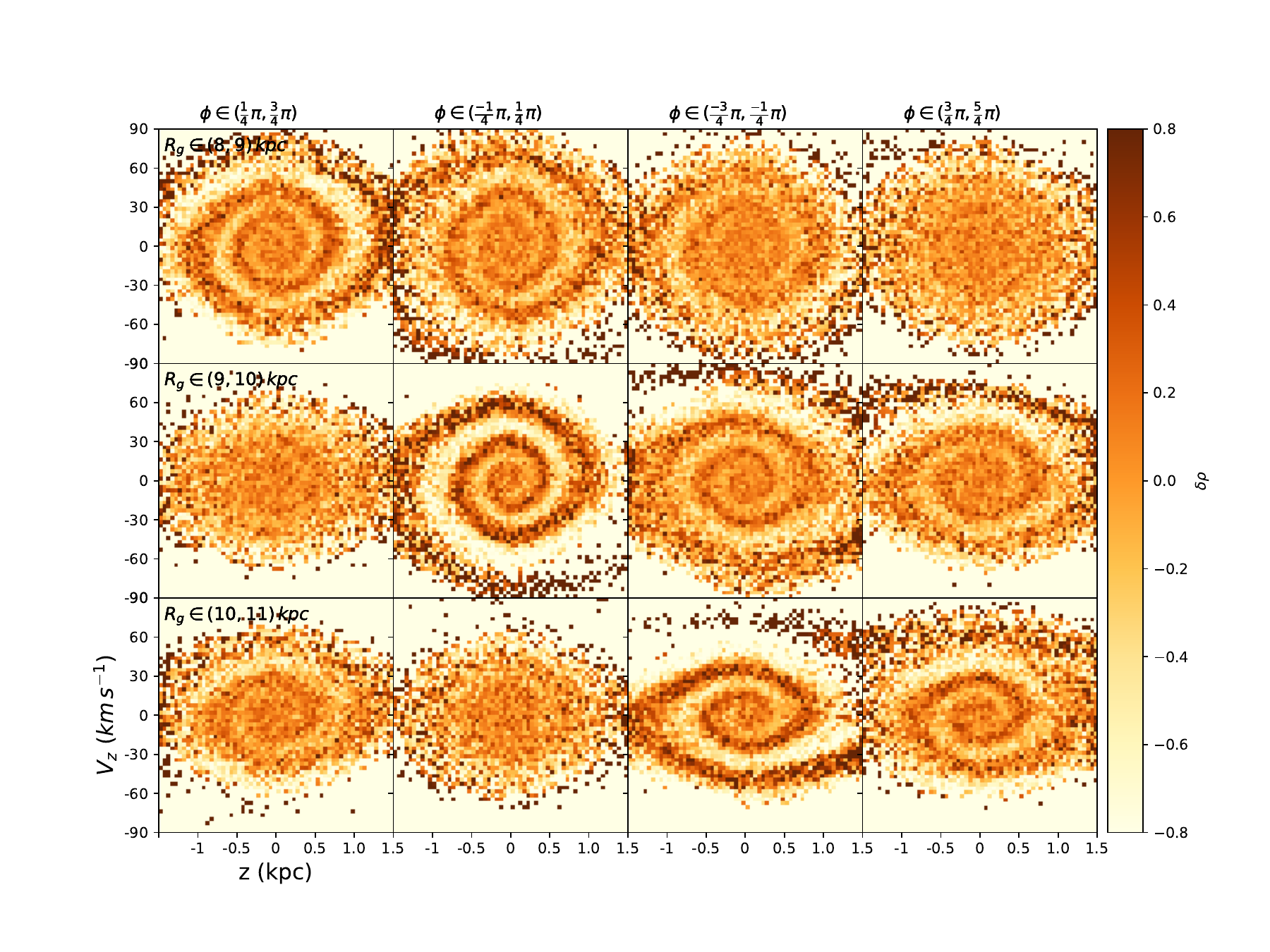}
    \caption{Phase spirals from the simulation color-coded by $\delta\rho$ at an evolution time of 0.5\,Gyr.
    %Left panel show the spirals from our first method with a vertical velocity kick of $\mathcal{N}(- 30,5)\,\rm km\,s^{-1}$. 
    The spirals are shown in different bins of $R_\mathrm{g}$ and azimuth from the simulation with a realistic velocity kick. From top to bottom, the three rows display phase spirals in bins of $R_\mathrm{g}$ of $8-9$, $9-10$ and $10-11$ kpc, respectively. The stars are binned by azimuth as indicated at the top of each column.}
    \label{simu_spiral}
\end{figure*}

\begin{figure}
    \centering
    \includegraphics[width=8cm, trim=0.6cm 0cm 1cm 1.5cm, clip]{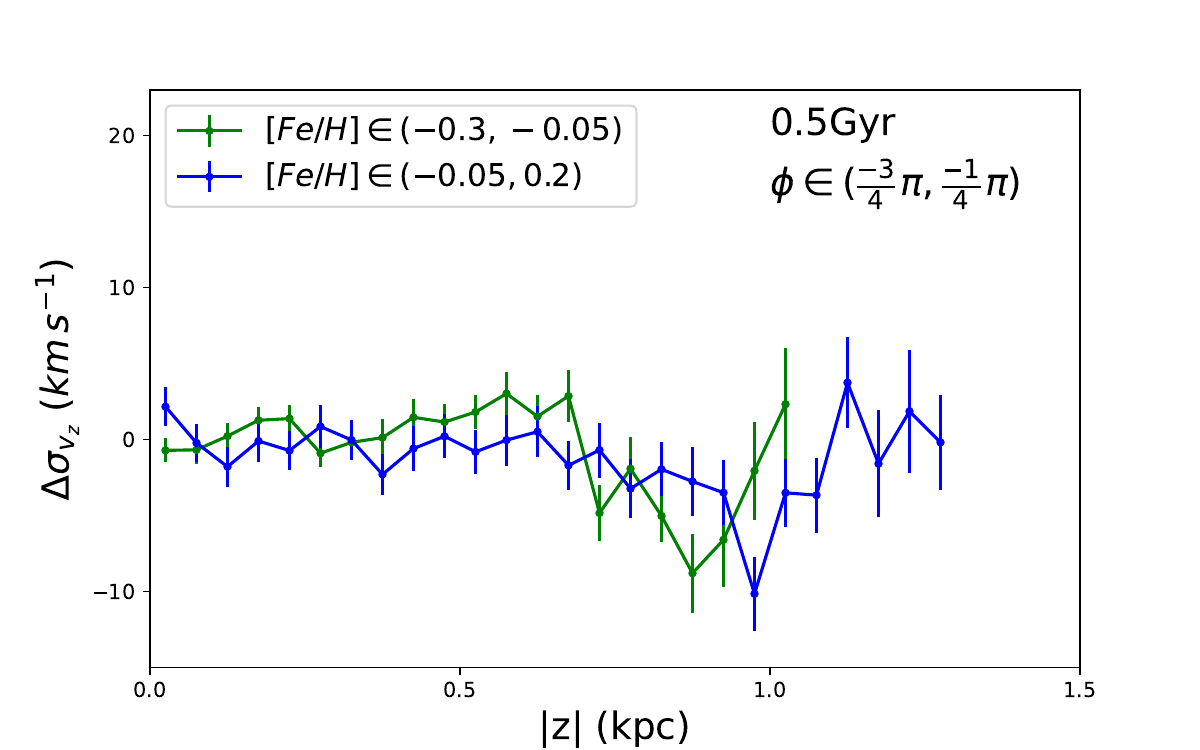}
    \caption{Dependency of the $\sigma_{v_z}$ NS asymmetry with [Fe/H] from simulation after a running time of 0.5\,Gyr. The particles from the Galactocentric radial range of $8\le R\le9$ kpc. We adopt an azimuthal criterion as indicated in the top of the panel. The metallicity [Fe/H] probed by each curve is indicated in the legend.}
    \label{feh_depend}
\end{figure}

\par\indent In order to mock the influence of the velocity perturbation on the distribution of metallicity, the test particles are labelled with a random [Fe/H] which varies with $R_\mathrm{g}$ and $z$ to mock a [Fe/H] distribution in $z-v_z$ phase space qualitatively  similar to observations \citep{2022arXiv220605534G}. According to the initial position of the particles in the $z-v_z$ phase space before perturbation, we give a random [Fe/H] value obeying a normal distribution with a mean of $\mu= -0.1(R_\mathrm{g}-8)-0.4 |z|$ and a dispersion of $\sigma= 0.1$ dex, where $R_\mathrm{g}$ and $z$ is in kpc. We integrate the particles under the Galactic potential for 0.6 Gyr until they form a stable $\mathrm{[Fe/H]}-z$ distribution, after which we introduce the vertical velocity disturbance.
%\subsection{Results from Simulations and discussion}
\subsection{NS asymmetries induced by the spiral}
Figure \ref{simu_3s} shows the evolution of the test particles in the radial range $9<R_\mathrm{g}<10$ kpc. The top row of Figure \ref{simu_3s} shows the contrast density map in phase space. The corresponding vertical velocity dispersion distribution and median [Fe/H] distribution are shown in the middle and bottom rows, respectively. We find peak and trough features similar to the observations. The intersection of the spiral and the $z$ axis corresponds to the decreases (troughs) in the velocity dispersion as indicated by the grey vertical dashed lines. The vertical velocity dispersion increases significantly at the $z$ region where the vertical velocity of the phase spiral component is high. We find a feature of the median metallicity profiles: the median [Fe/H] shows a bump feature at the height corresponding to the trough in the vertical velocity dispersion. Over time, the phase spiral becomes blurred, and the trough in the velocity dispersion curve and bump in the median metallicity profile become weaker. Note that we do not add background stars, thus the intensity of the NS asymmetries are enhanced in our simulations. Nevertheless, the qualitative result is not influenced.

\subsection{Shifts of the NS asymmetries found in the simulation}
We inspect the phase spirals from the simulation at the evolution time of 0.5\,Gyr, with the results shown in Figure \ref{dns_rho_feh} and \ref{simu_spiral}.  The right panels of Figure \ref{dns_rho_feh} display the asymmetries for particles with $\phi\in(-\frac{3\pi}{4} ,-\frac{\pi}{4})$, and show small shifts similar to the observations. 
\par The stars entering in a current spatial volume ($R-\phi$ volume) come from different locations of the Galactic plane, where they have in the past received quite different vertical velocity kicks at the time of the perturbation. The effect of mixing will make the phase spiral more disperse and thus change the peak locations and amplitudes of the $\Delta \sigma_{v_z}$ profiles.
\par By modelling the multiple impacts of a Sagittarius dwarf-like galaxy through a cold Milky Way-like disk using an $N$-body simulation, \cite{2021MNRAS.504.3168B} show that the $v_\phi$- and $ v_r$-traced phase spiral develops both in the clockwise and anticlockwise direction due to the disk’s shear. Our test particle simulation induces a velocity kick and, at an evolution time of 0.5\,Gyr, shows that the density-traced phase spiral emerges only in specific regions of azimuth. By binning along $R_\mathrm{g}$, our simulation shows that the shapes of the phase spirals wind tighter when the azimuth varies anticlockwise. Such varying shape of the phase spiral with azimuth has been shown in the Milky Way using the {\it Gaia} DR3 data \citep{Antoja2022}. The same limitation in $\phi$ for particles with different $R_\mathrm{g}$ will produce different epicyclic phase selection effects.  In the range of $\phi\in(-\frac{3\pi}{4},-\frac{\pi}{4})$, the spiral is more loosely wound in $8<R_\mathrm{g}<9$ kpc as compared to the more tightly wound spiral in the radial range of $10<R_\mathrm{g}<11$ kpc. The small shifts in the observations may partly result from these effects.
\par \cite{Binney2018} first point out that the evolution of the phase spiral depends on different vertical frequencies that are different along $R_\mathrm{g}$. This readily implies metallicity dependence of the phase spiral. By using a [Fe/H] label that only depends on $R_\mathrm{g}$, i.e. $\mu= -0.1(R_\mathrm{g}-8)$, we reproduce the metallicity dependence of $\Delta \sigma_{v_z}$ profiles. Figure \ref{feh_depend} shows the metallicity dependence from the simulation. The particles are in the $R$ range of $8-9$\,kpc at an evolution time of 0.5 Gyr. The result from the simulation shows a small shift similar to the observations, mainly due to the mixture effect from a radially and azimuthally dependent vertical perturbation.

\section{Discussion}
\label{discussion}
\subsection{Impact of selection effects}

 \par The primary selection effects in this work are firstly, those due to the sky coverage
of the cross-matched sample, and secondly, those propagated through from the spectroscopic survey, particularly on stellar colors, magnitudes and distances. Due to the location, survey strategy, and science goals of LAMOST, there is a strong focus on the anti-Galactic-Center direction and the Northern sky. We inspect the spatial distributions in the $R-z$ (and $R_\mathrm{g}-z$) space and the histograms of $R$ (and $R_\mathrm{g}$) of our samples, normalized by the total stellar numbers in the North and South respectively. There is only a significant difference in the $R$ distribution of the subsample in the range of $8<R<9$ kpc due to the survey footprint. The distributions form$R_\mathrm{g}$ are quite similar, even in the range of $8<R_\mathrm{g}<9$ kpc. From this we conclude that such selection effects do not affect the qualitative conclusions in our work.
  \par The complicated selection effects due to the spectral observations will significantly influence the calculation of the real number density profiles. However, in this work, we focus on the velocity dispersion profiles to study the variation in the shape of the phase spiral. The selection effects are trivial in the $v_z$ direction. Considering a narrow $z$ bin, we conclude the selection effects in this small bin are similar, and also are similar between the North and South. Thus, the selection effects do not greatly influence our results which are derived from the $\sigma_{z}$ profiles. In addition, we also check the density contrast profiles (1D) and maps (2D) in Figures \ref{dns_rho_feh} and \ref{spira_fs}, respectively. Though selection effects are clearly present, the Gaussian kernel smoothing helps alleviate such effects to some extent. The connection between the NS asymmetry and the variation in the shapes of the phase spirals are still clear in these two figures. Thus, we conclude that the selection effects do not influence our results and qualitative conclusions.

\subsection{Differences between simulation and observations}
\label{ssec_difference}
In this section we note the differences in the location of the peaks and troughs of the $\Delta \sigma_{v_z}$ profiles in the simulation compared to the observations and we consider several explanations. The shift between peaks is  larger in the simulation.
\par\indent The position of the peaks and shifts depend on the orientation of the spirals. The orientation of the spiral involves the initial phase angle of the perturbation, the wrapping speed (or frequency) at a certain radius, and the evolution time. As indicated by recent works \citep{Antoja2022,Frankel2022,Darragh2023}, the evolution time of spirals varies with different angular momenta ($R_\mathrm{g}$). This difference will also change the orientations of the spirals, and thus change the shifts of the peaks.
\par\indent Self-gravity might change the wrapping frequencies and the orientations of spirals. \cite{Darling2018} found that self-gravity can extend the duration of the phase spiral due to the persistent vertical oscillations. After about 1\,Gyr of evolution, the phase spiral without self-gravity winds tighter than a spiral with self-gravity. A subsequent theoretical work by \cite{Widrow2023} shows that self-gravity can change the shape of the phase spiral and slow down the phase mixing rate. The evolution time inferred with self-gravity would be shorter than that derived from simulations without self-gravity. The effect that the self-gravity slows down the phase mixing may change the orientation and shape of the phase spirals, and thus reduce the shift in the $\Delta \sigma_{v_z}$ profiles. However, the extent of such an influence due to self-gravity is still unclear, and may vary with different mechanisms of the phase spiral. The consideration of self-gravity is non-trival and is beyond the current scope of this work.

\subsection{Literature comparison}

\par The NS asymmetry in stellar number density was first discovered by \citet{2012ApJ...750L..41W} and later confirmed by more detailed investigations \citep{2013ApJ...777...91Y,2019MNRAS.482.1417B}. In these works, the number density profiles show excesses at $z \sim -0.4$ and $z \sim 0.7$ kpc, resulting in the trough and peak of the NS number density difference profile at $|z| \sim 0.4$ and $|z| \sim 0.7$ kpc, respectively. The locations of the peaks and troughs presented in our results are consistent with these works. \cite{2019ApJ...878L..31A} first found the NS asymmetry in the mean metallicity using the cross-matched sample of SDSS and {\it Gaia} DR2. The difference in the mean metallicity displays a wave-like oscillation, which coincides with the NS asymmetry in the stellar number density. \cite{2019ApJ...878L..31A} proposed that the NS asymmetries are induced by the phase spiral. Later, \cite{Guo2022} used a parameterized model to fit the $\sigma_{v_z}$ profiles of G/K dwarfs in the Solar Neighbourhood. The model is a superposition of a phase spiral similar to the observations and a smooth background assumed to be in dynamical equilibrium. They quantitatively explained the NS asymmetry in the $\sigma_{v_z}$ profiles by this phenomenological spiral model, which also gives a NS density difference profile consistent with the observations. Both studies show the strong connection between the NS asymmetry and the phase spiral.
\par Investigating the $z-v_z$ phase space, \cite{Bland-Hawthorn2019} dissected the sample according to the $\alpha$ abundance and the metallicity. They found that the phase spiral shows a clear trend in metallicity: the inner spiral is stronger for metal-rich stars, while the outer spiral is stronger for metal-poor stars, particularly in the $\alpha$-poor disk. They conclude that both the radial metallicity gradient and the age-velocity dispersion relation affect the metallicity dependence. \cite{2022arXiv220605534G} also found the metallicity dependence and the spiral feature in the vertical phase space color-coded by the residual metallicity, which they found to be especially obvious in the more distant radial bins.
\par Following the work of \cite{Guo2022}, we specifically investigate the NS asymmetry in $\sigma_{v_z}$ and the metallicity dependence utilizing the LAMOST data cross-matched with {\it Gaia} eDR3. We confirm again the connection between the NS asymmetry and the phase spiral through the coincidence in the NS asymmetry of $\sigma_{v_z}$, metallicity and the stellar number density. In addition, we focus more on the shift in the peak locations of $\Delta \sigma_{v_z}$ profiles. We relate this with the variation of the vertical potential at different radii, as the peak locations are actually the intersections between the phase snail and $z$ axis. We consider the metallicity dependence of the shift of the peak locations and the variation in the shapes of the phase spirals as due to the radial metallicity gradient and the variation of the vertical potential. The peaks and trough features provide an additional way to quantitatively measure the spiral intersections and thus the spiral shape. This will be helpful in relating the spiral shape with a certain vertical potential profile.

\section{Summary}
\label{summary}
\par\indent 
\par Using thin disk stars selected from LAMOST and {\it Gaia} for which we have full 6D information, as well as [Fe/H], we inspect the NS asymmetry of the vertical velocity dispersion in the Galactocentric radial range and guiding center radial range of $8-11$\,kpc. We find peak and trough features in $\sigma_{v_z}-z$ profiles that appear at all radial ranges and in subsamples separated by [Fe/H]. We demonstrate that the peaks and troughs in the $\Delta \sigma_{v_z}$ profiles are indicators of the intersections between the phase spiral and $z$ axis. The  difference in $\sigma_{v_z}$ between the North and South, i.e., the $\Delta \sigma_{v_z}-z$ NS asymmetry profile, shows a wave-like pattern. We find strong correlation between the $\Delta \sigma_{v_z}-z$ NS asymmetry and number density NS asymmetry, which supports their origin from the phase spiral. On the other hand, we find a weak wave-like signal that is related to the phase spiral from the [Fe/H] NS asymmetry profile near the solar radius.

\par We also find a variation of the peak locations in $\Delta \sigma_{v_z}$ profiles binned by Galactocentric radius $R$ and by guiding center radius $R_\mathrm{g}$. The peak locations of stars with smaller $R$ ($R_\mathrm{g}$) are shifted to larger values of $|z|$. This trend becomes even more obvious when binning the sample with $R_\mathrm{g}$. This shift in the peak locations reflects the variation in the shape of the phase spiral, which intrinsically indicates the vertical potential at different radii. This shift also exists in the subsamples separated by metallicity, reflecting the influence of the radial metallicity gradient. We separate the total sample into three subsamples according only to metallicity, and roughly measure the shapes of the three phase spirals. The spiral of the metal-rich sample is located on the outward side along the metal-poor spiral. The metal-rich stars have smaller guiding center radii and thus a steeper vertical potential, resulting in the spiral lying along the outward side of the metal-poor phase space spiral. The result from the 2D spiral shape is consistent with the results from the 1D $\Delta \sigma_{v_z}$ profiles.

\par We perform a test particle simulation similar to that of \cite{Binney2018} in order to qualitatively recover our results found in the observations. Comparable to the observations, we find the wave-like oscillations in the $\Delta \sigma_{v_z}$ profiles and the shift in the peak locations for the simulated samples in different radial and chemical ranges. The vertical velocity perturbation can excite the metal-rich stars to higher regions from the disk and produce a metallicity bump relative to the background, which then results in the NS asymmetry in metallicity. The radial and azimuthal variations of the velocity kick and azimuth limitation produce a different epicyclic selection effect for the samples separated into different bins of $R_\mathrm{g}$. This selection effect along with the mixing of the stars are likely the main causes of the small amplitude and shift seen in the observations.

\par Our results from the simulation are only qualitatively similar to the observations. Differences remain between the simulation and observations. 
The spiral shape is influenced by the initial phase angle after the perturbation, the wrapping speed at each radius, and the evolution time since the perturbation. These are found to vary with the angular momentum and the orbital properties of the stars as shown in recent works \citep{Antoja2022,Frankel2022,Darragh2023}. 
A quantitative explanation of the NS asymmetry in metallicity needs more complicated modelling and quantitative consideration of the relation between the spiral shape and the vertical potential. This will be analysed in future work (Guo et al. in preparation), as well as the presentation of carefully constructed perturbation modelling.% Our study helps shed light on the possibility to use the phase spiral as a means to study the radial metallicity gradient and the migration induced by the recent perturbation.

%==============================================================
\section*{Acknowledgements}
We thank Feng Wang, Qiang Yuan, and Zhaoyu Li for helpful discussions and acknowledge the National Natural Science Foundation of China (NSFC) under grant Nos. 12373033, 12303023, 11873034, and 12011530421, the Department of Science and Technology of Hubei Province for the Outstanding Youth Fund (2019CFA087), the Cultivation Project for LAMOST Scientific Payoff and Research Achievement of CAMS-CAS, and the science research grants from the China Manned Space Project including the CSST Milky Way and Nearby Galaxies Survey on Dust and Extinction Project CMS-CSST-2021-A09 and No. CMS-CSST-2021-A08. RG is supported by Initiative Postdocs Supporting Program (No. BX2021183), funded by China Postdoctoral Science Foundation. This work has made use of data from the European Space Agency (ESA) mission {\it Gaia} (https://www.cosmos.esa.int/gaia), processed by the {\it Gaia} Data Processing and Analysis Consortium (DPAC, https://www.cosmos.esa.int/web/gaia/dpac/consortium). Funding for the DPAC has been provided by national institutions, in particular the institutions participating in the {\it Gaia} Multilateral Agreement. Guoshoujing Telescope (LAMOST) is a National Major Scientific Project built by the Chinese Academy of Sciences. Funding for the project has been provided by the National Development and Reform Commission. LAMOST is operated and managed by the National Astronomical Observatories, Chinese Academy of Sciences.
%==============================================================

\section*{Data Availability}
The data supporting this article will be shared upon reasonable request sent to the corresponding authors.

\bibliographystyle{mnras}
\bibliography{main}

\end{document}